\documentclass[twocolumn,twocolappendix]{aastex631}

\usepackage{stackengine}
\newsavebox\mybox

\newcommand\hi{H~\textsc{i}}
\newcommand\lb{Local Bubble}

\newcommand{\nsideof}[1]{$N_{\rm side}~=~{#1}$}
\newcommand{\eqref}[1]{Equation~\ref{#1}}
\newcommand{\secref}[1]{Section~\ref{#1}}
\newcommand{\figref}[1]{Figure~\ref{#1}}
\newcommand{\tabref}[1]{Table~\ref{#1}}
\newcommand{\eq}[1]{\begin{equation} #1 \end{equation}} 
\shorttitle{Imprints of the Local Bubble and Dust Complexity on Polarized Dust Emission}
\shortauthors{Halal et al.}

\begin{document}

\title{Imprints of the Local Bubble and Dust Complexity on Polarized Dust Emission}

\correspondingauthor{George Halal}
\email{georgech@stanford.edu}

\author[0000-0003-2221-3018]{George Halal}
\affiliation{Department of Physics, Stanford University, Stanford, CA 94305, USA}
\affiliation{Kavli Institute for Particle Astrophysics and Cosmology, Stanford, CA 94305, USA}

\author[0000-0002-7633-3376]{S. E. Clark}
\affiliation{Department of Physics, Stanford University, Stanford, CA 94305, USA}
\affiliation{Kavli Institute for Particle Astrophysics and Cosmology, Stanford, CA 94305, USA}

\author[0000-0001-8749-1436]{Mehrnoosh Tahani}
\affiliation{Department of Physics, Stanford University, Stanford, CA 94305, USA}
\affiliation{Kavli Institute for Particle Astrophysics and Cosmology, Stanford, CA 94305, USA}
\affiliation{Banting and KIPAC Fellow, Stanford, CA 94305, USA}



\begin{abstract} 
Using~3D dust maps and Planck polarized dust emission data, we investigate the influence of the~3D geometry of the nearby interstellar medium~(ISM) on the statistics of the dust polarization on large~($80'$) scales. We test recent models that assume that the magnetic field probed by the polarized dust emission is preferentially tangential to the Local Bubble wall, but we do not find an imprint of the Local Bubble geometry on the dust polarization fraction. We also test the hypothesis that the complexity of the~3D dust distribution drives some of the measured variation of the dust polarization fraction. We compare sightlines with similar total column densities and find that, on average, the dust polarization fraction decreases when the dust column is substantially distributed among multiple components at different distances. Conversely, the dust polarization fraction is higher for sightlines where the dust is more concentrated in~3D space. This finding is statistically significant for the dust within~1.25~kpc, but the effect disappears if we only consider dust within~270~pc. In conclusion, we find that the extended~3D dust distribution, rather than solely the dust associated with the Local Bubble, plays a role in determining the observed dust polarization fraction at~80$'$. This conclusion is consistent with a simple analytical prediction and remains robust under various modifications to the analysis. These results illuminate the relationship between the~3D geometry of the ISM and tracers of the interstellar magnetic field. We discuss implications for our understanding of the polarized dust foreground to the cosmic microwave background. 
\end{abstract} 



\section{Introduction} \label{sec:intro}
Interstellar magnetic fields play an important role in various astrophysical processes \citep[see, e.g.,][]{Ferriere:2001, Heiles:2012, Pattle:2023}. However, little is known about the magnetic field structure in the nearby interstellar medium (ISM). Some works have suggested connections between the local magnetic field structure and other tracers of ISM morphology, perhaps due to dynamical influences, e.g. the formation of structures like superbubbles \citep[e.g.,][]{Santos:2011, Frisch:2012, Berdyugin:2014, Tahani:2022a, Tahani:2022b}. 

Aspherical dust grains in the ISM emit photons with an electric field oriented preferentially along their long axes \citep{1975duun.book..155P}. The short axes of typical dust grains are preferentially aligned with the local magnetic field orientation \citep{Andersson:2015}. As a result, their thermal emission is partially polarized perpendicular to the orientation of the magnetic field. Therefore, measurements of the polarized dust emission are used as a probe for the plane-of-sky magnetic field orientation in dusty regions of the ISM.

Recent evidence suggests that variations in the fractional polarization of the dust emission, i.e., the ratio of the polarized to the total intensity of the dust emission, over large angular scales in the diffuse sky are mainly driven by the structure of the magnetic field. Henceforth, we will refer to the fractional polarization of the dust emission as the dust polarization fraction. \citet{PlanckCollaboration:2020} probed the influence of the magnetic field geometry on the dust polarization fraction by comparing the local polarization angle dispersion with the~353~GHz polarization fraction. They calculated the polarization angle dispersion for an annulus with inner and outer radii of~40$'$ and~120$'$, and found that the~353~GHz polarization fraction is anti-correlated with the local polarization angle dispersion at~160$'$ resolution. They showed that this relationship is consistent with models that only include topological effects of the turbulent magnetic field, but otherwise have uniform dust properties and alignment. \citet{PlanckCollaboration:2020} conclude that the dust polarization fraction and the dispersion of polarization angles are similarly sensitive to the structure of the magnetic field. \citet{Hensley:2019} further showed that some of the variability in the dispersion of polarization angles, and thus the dust polarization fraction, can be explained by the magnetic inclination angle, i.e., the angle between the magnetic field and the plane of the sky. The dust polarization fraction is maximized when the magnetic field is tangential to the plane of the sky and zero when it is parallel to the line of sight. \citet{Chen:2019} and \citet{Sullivan:2021} used statistical properties of the observed dust polarization fraction to estimate the average inclination angle of molecular cloud-scale magnetic fields. 

The observed dust polarization fraction also depends on other factors, such as the dust grain alignment efficiency \citep{King:2019, Medan:2019}, the phase distribution of the neutral interstellar medium \citep{Lei:2023}, and measurement noise. However, the~3D structure of the magnetic field is one of the major factors \citep{Clark:2018, Hensley:2019, PlanckCollaboration:2020}. 

Our Sun's current location is near the center of a superbubble, which is thought to have been created by supernova explosions within the past 10-15$~\times~10^6$ years \citep{1987ARA&A..25..303C, Maiz-Apellaniz:2001, Breitschwerdt:2016, Schulreich:2023}. It is commonly known as the \lb, Local Cavity, or Local Chimney \citep{Welsh:2004, Puspitarini:2012}. Since star formation tends to be concentrated, sequential supernovae are common \citep{Zucker:2022,Watkins:2023,Barnes:2023,Sandstrom:2023}. Supernova explosions sweep up matter and magnetic field lines, leaving behind low-density superbubbles on the order of hundreds of parsecs in diameter \citep{Kim:2015}. The swept-up matter is compressed into a shell surrounding the expanding superbubble, which is thought to trigger the formation of dense gas and stars \citep{Elmegreen:2011, Dawson:2013, Inutsuka:2015}. 

Thus, it is reasonable to expect that the formation of the \lb\ dramatically influenced the magnetic field geometry in the nearby ISM. Some studies have aimed at modeling the geometry of the wall of cold neutral gas and dust surrounding the \lb\ \citep[e.g.,][]{Alves:2018, Pelgrims:2020}. Since the geometry of the magnetic field affects the measured dust polarization fraction, we search for an imprint of the \lb\ geometry on the dust polarization fraction in this paper.

Additionally, one of the probes of the~3D spatial distribution of the neutral ISM is dust extinction toward stars. This is due to the scattering and absorption of starlight by dust. The extinction of a star's apparent magnitude is correlated with the dust column density along the line of sight from the observer to that star. The Gaia survey provided accurate distances to more than a billion stars within a few kiloparsecs from the Sun. Combining this distance information with the level of extinction towards each star has been transformative for the construction of~3D maps of the differential dust extinction \citep{Lallement:2019, Leike:2020, Vergely:2022, Edenhofer:2023}. We use several~3D dust maps to quantify the complexity of the spatial distribution of the dust along the line of sight. We use that to explore the relationship between the~3D dust distribution and the measured dust polarization fraction.

In this work, we investigate the relationship between the 3D geometry of the nearby ISM and the dust polarization fraction. We start by introducing the data we use in \secref{sec:data}. In \secref{sec:LB}, we search for an imprint of the \lb\ geometry on the dust polarization fraction. In \secref{sec:3ddust}, we test how the dust polarization fraction is affected by the line-of-sight complexity of the dust. We discuss the implications of our results and conclude in \secref{sec:conc}.

\section{Data} \label{sec:data}

\begin{figure}[t!]
\includegraphics[width=\columnwidth]{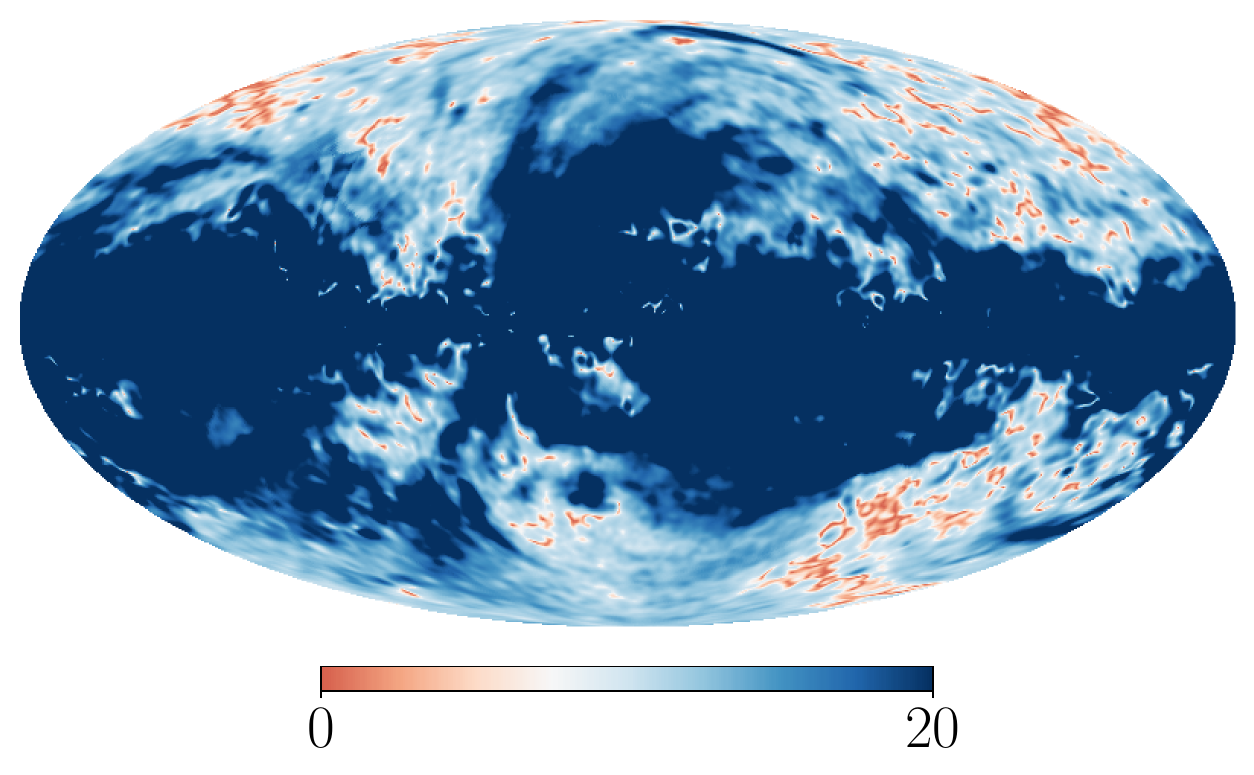}
\caption{A map of the debiased signal-to-noise ratio of the Planck GNILC polarization fraction at~$80'$. This is plotted with a diverging linear colorbar centered on~3, the cutoff we use as part of our sightline selections in Sections~\ref{sec:LB}~and~\ref{sec:3ddust}, with the allowed regions shown in blue.
\label{fig:snr}}
\end{figure}

\begin{figure*}[t!]
\includegraphics[width=2\columnwidth]{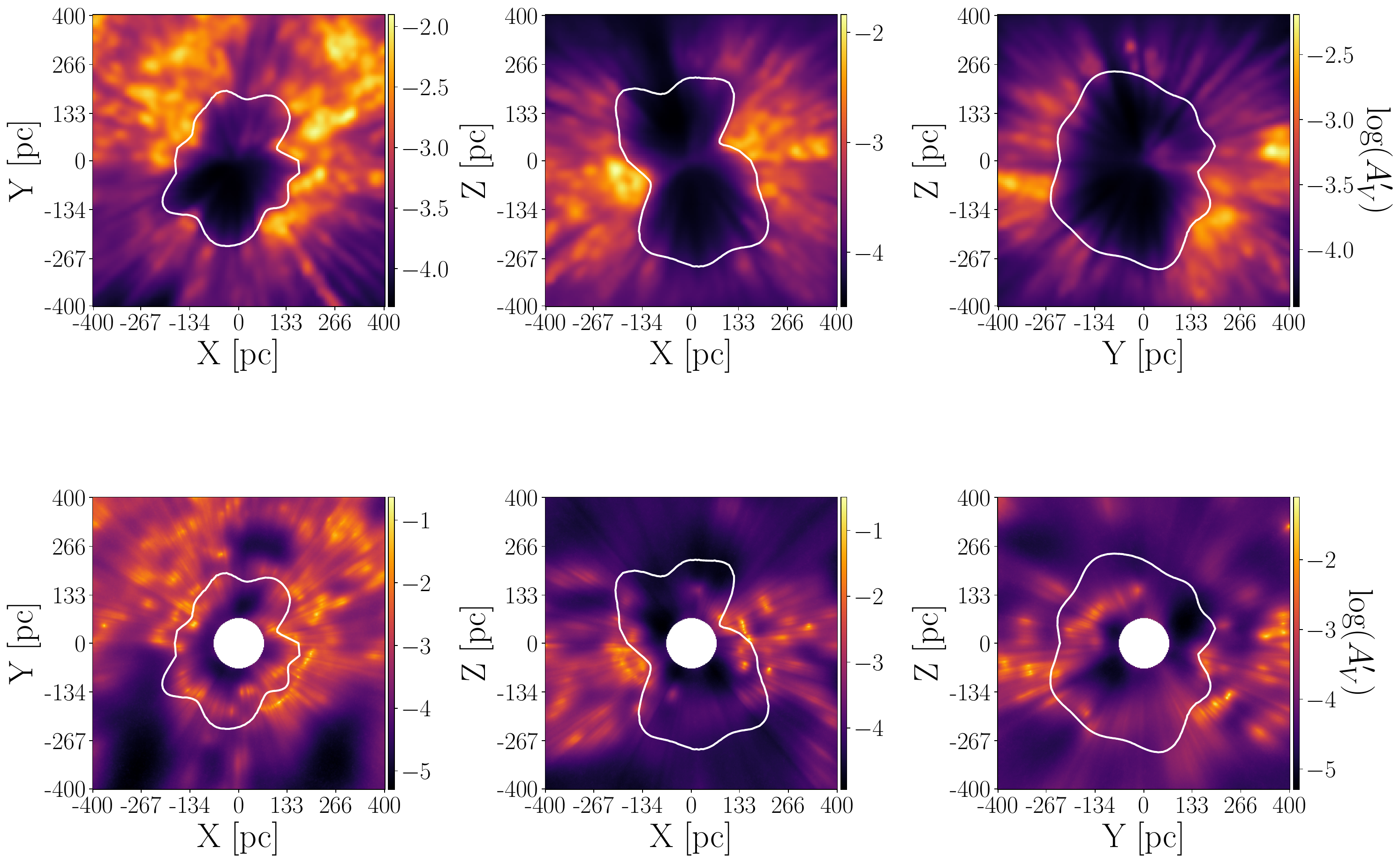}
\caption{Slices through the XY (left), XZ (middle), and YZ (right) planes of the~3D reconstructed differential extinction maps of \citet{Lallement:2019} (top) and \citet{Edenhofer:2023} (bottom). The Sun is at the origin. The positive X axis points towards the Galactic center at~$l~=~0^\circ$, the positive Y axis points towards~$l~=~90^\circ$ in the Galactic plane, and the positive Z axis points out of the plane in the direction of the Galactic North pole. The slices show the log of the differential extinction, which is in units of magnitudes per parsec. The subpanels only extend up to~400~pc in each direction for a direct comparison. The model for the \lb\ surface geometry of \citet{Pelgrims:2020} is overplotted in white in each subpanel. The white disk at the center of the bottom panel is due to missing data within~70~pc of the Sun in the \citet{Edenhofer:2023} maps.
\label{fig:slices}}
\end{figure*}

\subsection{Planck Data Products} \label{subsec:red}
We use the~80$'$~R3.00 Planck data processed with the Generalized Needlet Internal Linear Combination \citep[GNILC;][]{2011MNRAS.418..467R} method at~353 GHz to remove the Cosmic Infrared Background (CIB) radiation from the Galactic dust emission \citep{PlanckCollaboration:2016}. Following the fiducial offset corrections adopted by the Planck collaboration, we subtract~452~$\mu{\rm K}_{\rm CMB}$ from the GNILC total intensity map to correct for the CIB monopole then add a Galactic offset correction of~63~$\mu{\rm K}_{\rm CMB}$ \citep{PlanckCollaboration:2020}. Because the GNILC data are at FWHM=~$80'$, we downgrade the maps from their native HEALPix pixelization \citep{2005ApJ...622..759G} at~\nsideof{2048} to~\nsideof{64}. We also use the R3.01 Planck data at~353 GHz, smoothed to~80$'$ as a cross-check \citep{PlanckCollaboration:2020b}. All of these maps use the COSMO polarization convention. We do not convert to the IAU polarization convention.

We use the modified asymptotic estimator of \citet{Plaszczynski:2014} to debias the polarized intensity of the GNILC dust emission map and the associated uncertainty. We obtain a signal-to-noise ratio map of the dust polarization fraction,~SNR$_p$, shown in \figref{fig:snr}.

We estimate the total dust extinction over the full sky using Planck data products. \citet{PlanckCollaboration:2014} and \citet{PlanckCollaboration:2016} fit a modified blackbody spectrum to the GNILC dust maps at different frequencies to estimate the dust temperature, spectral index, and optical depth over the sky. The dust optical depth is correlated with the reddening of quasars \citep{PlanckCollaboration:2014}. Using this observation, \citet{PlanckCollaboration:2016} multiply the GNILC dust optical depth map by a factor of~$1.49~\times~10^{-4}$~mag to construct a GNILC~$E(B-V)$ map. We query the publicly available \texttt{dustmaps} Python package \citep{Green2018} for the~\citet{PlanckCollaboration:2016} GNILC~$E(B-V)$ map. Assuming a standard extinction law, we multiply the GNILC~$E(B-V)$ map by~3.1 to obtain~$A_{V}^{\rm Planck}$.

\subsection{3D Dust Maps and Local Bubble Geometries} \label{subsec:dustdata}

In \secref{sec:LB}, we use the~3D model of the \lb\ surface geometry constructed by \citet{Pelgrims:2020}. To create this model, \citet{Pelgrims:2020} extract distances to the first high dust density regions around the Sun from  the~3D Cartesian map of dust differential extinction constructed by \citet{Lallement:2019}. They smooth the map of the distances to the \lb\ surface by filtering out spherical harmonic modes above some threshold to remove small-scale fluctuations that might appear due to an inhomogeneous distribution of the dust density on the small scales. In this work, we use the map filtered to $\ell_{\rm max}=10$, which is the map used in their analysis. The \citet{Lallement:2019} dust map is based on data from Gaia DR2 \citep{GaiaCollaboration:2018} and 2MASS \citep{Skrutskie:2006} and spans~$6~\times~6~\times~0.8~{\rm kpc}^3$ in the Heliocentric right-handed Galactic-XYZ coordinates. It has a voxel volume of~125~pc$^3$ and a spatial resolution of~25~pc. Example slices of this map with the \citet{Pelgrims:2020} model overplotted are shown in \figref{fig:slices}. 

We also use the~3D model of the \lb\ surface constructed by \citet{O'Neill:2024} for a brief investigation in \secref{sec:LB}. They use the~3D dust differential extinction provided by \citet{Edenhofer:2023} to extract the distance to the \lb\ in all directions as a region of higher dust density around the Sun. 

In \secref{sec:3ddust}, we query the~12 posterior samples for the~3D dust maps provided by \citet{Edenhofer:2023} via the publicly available \texttt{dustmaps} Python package \citep{Green2018} at their plane-of-sky native angular resolution of~$14'$, which corresponds to a HEALPix pixelization scheme at~\nsideof{256}. These~3D dust maps leverage distance and extinction estimates to stars from \citet{Zhang:2023}, which are derived from Gaia DR3 data \citep{GaiaCollaboration:2023}. The distance resolution of these maps varies from~0.4~pc at~69~pc to~7~pc at~1.25~kpc. We query the map using uniform distance bins of~7~pc. The map is provided in unitless extinction values defined in \citet{Zhang:2023}. We multiply the map by a factor of~2.8 to convert it to Johnson’s~$V$-band~$A'_V$ \citep{Zhang:2023}. We then convert~$A'_V$ to volume density of hydrogen nuclei~($n_{\rm H}$) using the extinction curve from \citet{Fitzpatrick:2019} to convert~$A_G~=~0.796~A_V$ and the relationship~$A_G~/~N_{\rm H}~=~4~\times~10^{-22}~{\rm cm}^2~{\rm mag}$ from \cite{Zucker:2021} and \citet{Bialy:2021}. To match the resolution of the dust polarization fraction and smooth out small-scale fluctuations in the map, we smooth the HEALPix sphere at each distance bin to a FWHM=~$80'$ then repixelate it to~\nsideof{64}. Example slices of the raw map of the mean of the posterior samples are also shown in \figref{fig:slices} with the \citet{Pelgrims:2020} model overplotted for comparison.

In \secref{sec:3ddust}, we also make comparisons with the~3D dust maps provided by \citet{Leike:2020}. These maps leverage distance and extinction estimates from the StarHorse catalog \citep{Anders:2019}, which combines data from Gaia~DR2 \citep{GaiaCollaboration:2018}, ALLWISE \citep{Cutri:2013}, PANSTARRS \citep{Flewelling:2020}, and~2MASS \citep{Skrutskie:2006}. The \citet{Leike:2020} maps span~$740~\times~740~\times~540~{\rm pc}^3$ in the Heliocentric Galactic-XYZ coordinates, respectively, with a voxel size of~1~pc$^3$ and spatial resolution of~1~pc. We also query the~12 posterior samples of the \citet{Leike:2020} maps via the \texttt{dustmaps} package. The \citet{Leike:2020} maps are given in optical depth in the Gaia~G band per~1~pc. We convert to~$n_{\rm H}$ following \cite{Zucker:2021} and \citet{Bialy:2021}.

\subsection{Galactic Faraday Rotation Measure} \label{subsec:rotmeas} 
We use the all-sky Galactic Faraday rotation measure (RM) map produced by \citet{Hutschenreuter:2023} for a brief investigation in \secref{sec:LB}. Using information field theory, a Bayesian inference framework for fields \citep{Ensslin:2019}, \citet{Hutschenreuter:2023} disentangle the Galactic contribution to the RM from the compiled RM catalogs of polarized radio sources such as radio galaxies \citep{VanEck:2023}, supplemented by Galactic pulsar dispersion measures \citep{Manchester:2005}, as well as data on Galactic bremsstrahlung emission \citep{PlanckCollaboration:2016b} and the hydrogen~$\alpha$ spectral line \citep{Finkbeiner:2003}.

\section{No Imprint of the Local Bubble on the Dust Polarization Fraction} \label{sec:LB}

\subsection{Motivation}
\label{subsec:assump}
We begin the exploration of the effect of different geometrical factors on the dust polarization fraction by searching for an imprint of the geometry of the dust wall surrounding the \lb\ on the Planck~353~GHz dust polarization fraction. In this subsection, we discuss the assumptions made in previous studies regarding the \lb\ surface. These assumptions help us design a test for studying this effect in the next subsection. A significant detection of an imprint of the \lb\ geometry on the dust polarization fraction would validate these assumptions.

The first assumption is that the observed polarized dust emission is dominated by dust in the \lb\ wall at the relevant angular scales and Galactic latitudes. This assumption, made in several analyses \citep[e.g.,][]{Alves:2018, Pelgrims:2020, ONeill:2023}, is supported by several studies using optical starlight polarization data \citep{Leroy:1999,Andersson:2006,Santos:2011,Frisch:2015,Medan:2019,Cotton:2019,Skalidis:2019}. The alignment of neutral hydrogen structures at local velocities with starlight polarization toward stars at distances within a few hundred parsecs is consistent with a picture where most of these structures are positioned at comparable distances within the \lb\ at high Galactic latitudes~($|b|~>~30^\circ$) \citep{Clark:2014}. \citet{Gontcharov:2019} observed that starlight polarization fraction plateaus after~150-250~pc across the sky. By comparing the~353~GHz polarized emission with the polarized optical starlight, \citet{Skalidis:2019} find that most of the~353~GHz polarized emission signal is captured within the first~250~pc at~$|b|~>~60^\circ$, suggesting the presence of a dust wall around that distance. They, however, find that this conclusion does not hold at~$30^\circ~<~|b|~<~60^\circ$.

The second assumption is that the magnetic field’s inclination is tangential to the surface of the \lb, which stems from the model assumed for the formation of the \lb. To fit a model of the \lb\ magnetic field, \citet{Alves:2018} and \citet{Pelgrims:2020} assume that all the swept-up matter and field lines due to the supernova explosions that formed the \lb\ are squeezed into a thin layer that follows its surface, leading the magnetic field lines to be tangent to the surface. \citet{ONeill:2023} also make this assumption to project the observed polarization angles of the dust emission onto the Bubble's surface and build a~3D model of the Bubble wall magnetic field. 

Other works have found magnetic field structure tangential to bubbles on supernova scales, observationally \citep[e.g.,][]{Kothes:2009, West:2016, Tahani:2022a, Tahani:2022b} and in simulations \citep{Kim:2015, Maconi:2023}. It has also been shown on the scales of H\textsc{ii} regions, observationally \citep{Tahani:2023} and in simulations \citep{Krumholz:2007}. On the scale of superbubbles, such as our \lb, a comparison of the plane-of-the-sky and line-of-sight magnetic field strengths as well as measurements of the dust polarization fraction towards regions associated with the Orion-Eridanus superbubble suggest that the large-scale magnetic field in the region was primarily shaped by the expanding superbubble and is tangential to its surface \citep{Heiles:1997, Soler:2018}. However, there is no direct evidence that the nearby magnetic field is preferentially tangential to the \lb\ surface.

The shape of the \lb\ wall has been modeled differently in different works. Some studies fit a generalized parametric geometry, such as an ellipsoid \citep{Alves:2018}, in an attempt to fit general properties of the \lb\ magnetic field on large scales. Others model the detailed boundary of the \lb\ using~3D maps of the dust extinction \citep{Pelgrims:2020, O'Neill:2024}. These models therefore vary based on both the variations in the different~3D dust maps used as well as the methodology applied to these maps to define a \lb\ surface. \citet{Pelgrims:2020} model the radial distance of the \lb\ wall from the Sun in each direction as the first distance where the second derivative with respect to the distance of the differential extinction constructed by \citet{Lallement:2019} reaches zero, i.e., the first inflection point,~${\rm d}^2 A_{V}'(r)/{\rm d}r^2~=~0$. \citet{ONeill:2023} and \citet{Zucker:2022} use the geometry defined by \citet{Pelgrims:2020} in their work. \citet{O'Neill:2024} employ a similar methodology to \citet{Pelgrims:2020}, using the differential extinction maps constructed by \citet{Edenhofer:2023} instead to construct their model. Other studies use different tracers to model the geometry of the Bubble. \citet{2017ApJ...834...33L} assumes that the measured X-ray intensity is proportional to the distance to the Bubble in the considered direction. Several other tracers such as~NaI absorption measurements \citep{Sfeir:1999,Lallement:2003}, stellar color excess measurements \citep{Lallement:2014}, and diffuse interstellar bands \citep{Farhang:2019}, have also been used for constructing models of the Bubble wall geometry. These geometries vary significantly from one to another. While some model the \lb\ as a closed surface \citep{Pelgrims:2020}, others describe the same structure as a Local Chimney, i.e., open in one or both directions away from the disk and funneling material into the Milky Way's halo \citep[e.g.,][]{Heiles:1984, Sfeir:1999, Lallement:2003, Marchal:2023, O'Neill:2024}. Also, the \lb\ surface may have tunnels to surrounding cavities like the Gum Nebula and/or GSH238+00+09 rather than having a closed geometry \citep[e.g.,][]{Welsh:1991, Lallement:2003, Marchal:2023, O'Neill:2024}. 

We aim to test whether the degree-scale structure of the \lb\ is measurably imprinted in the statistics of polarized dust emission. We use the most recent models of the \lb\ geometry constructed based on~3D dust mapping. For our main analysis, we use the \citet{Pelgrims:2020} model, which has a closed geometry. However, we also perform a brief test using the \citet{O'Neill:2024} model, whose geometry is not fully connected. In each case, we test whether a magnetic field that is tangential to the \lb\ wall is a statistical driver of the observed variation in the dust polarization fraction.

\subsection{Testing the Dependence of the Dust Polarization Fraction on the Magnetic Inclination Angle} 
\label{subsec:LBtest}

\begin{figure}[t!]
\includegraphics[width=\columnwidth]{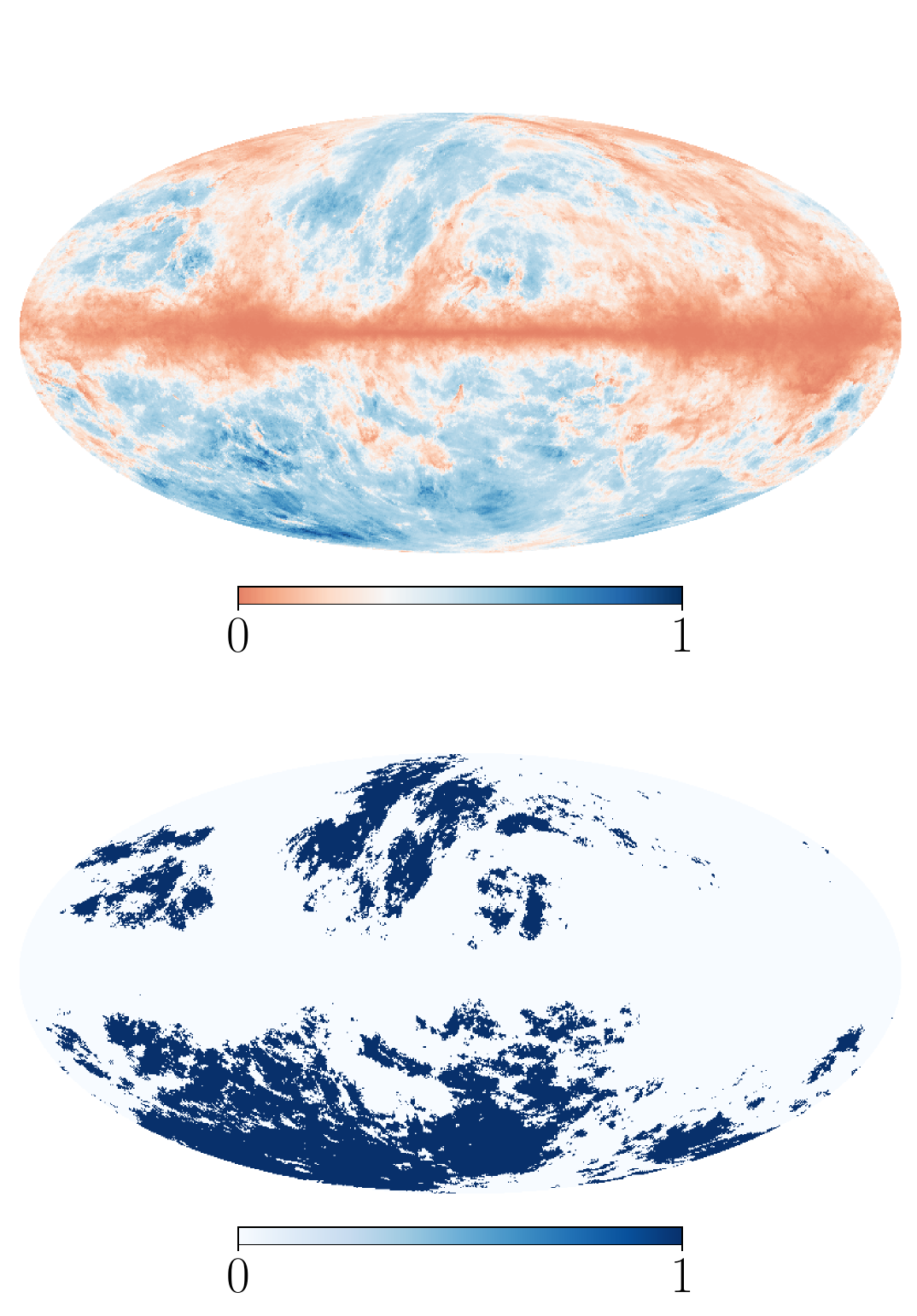}
\caption{Maps used for selecting sightlines for the analysis described in \secref{sec:LB}. \textit{Top panel:} A map of the ratio of the \citet{Lallement:2019}~3D dust differential extinction map integrated within~50~pc of the \lb\ surface defined by \citet{Pelgrims:2020}, $A_{V}^{\rm LB}$, over the Planck dust extinction, $A_{V}^{\rm Planck}$. This is plotted with a diverging colorbar centered on the~75th percentile (0.2), the cutoff we use in our sightline selection, with the allowed regions shown in blue. \textit{Bottom panel:} A map of the mask of the selected region, combining $A_{V}^{\rm LB} / A_{V}^{\rm Planck}>0.2$ (top panel) with~SNR$_p>3$ (\figref{fig:snr}).
\label{fig:masks_LB}}
\end{figure}

Using~3D dust extinction maps and a model of the \lb\ surface geometry, we can measure the angle between the line of sight and the local magnetic field orientation projected onto the Bubble's surface in each direction. If the magnetic field lines were tangential to the \lb\ surface as discussed in \secref{subsec:assump}, the angle between the line of sight and the Bubble's surface should on average be correlated with the measured dust polarization fraction along sightlines where the dust is concentrated in the \lb\ wall. Therefore, by quantifying the correlation between this angle and the dust polarization fraction for different sightlines, we can test whether there is measurable evidence that the magnetic field traced by the dust polarization is preferentially tangential to the \lb\ surface.

The magnetic inclination angle~$\gamma$, i.e., the angle between the magnetic field orientation and the plane of the sky, affects the measured dust polarization fraction. The polarization fraction,~$p$, is defined as
\eq{p~=~P/I~=~\sqrt{Q^2 + U^2}/I,}
where~$P$ is the debiased polarized intensity as described in \secref{sec:data},~$Q$ and~$U$ are the Stokes parameters, and~$I$ is the total unpolarized intensity. Assuming uniform grain properties, the Stokes~$I$,~$Q$, and~$U$ parameters of the dust emission can be written as \citep{Fiege:2000, Padoan:2001, Pelkonen:2007}
\eq{I~=\int \epsilon \rho \,ds - \frac{1}{2} \int \alpha \epsilon \rho \left(\cos^2{\gamma} - \frac{2}{3}\right)\,ds, \label{eq:I}}
\eq{Q~= -\int \alpha \epsilon \rho \cos{2\psi}\cos^2{\gamma} \,ds, \label{eq:Q}}
\eq{U~= -\int \alpha \epsilon \rho \sin{2\psi}\cos^2{\gamma} \,ds, \label{eq:U}}
where~$\rho$ is the volume density,~$ds$ is a distance segment along the line of sight,~$\epsilon$ is the dust emissivity,~$\alpha$ is a coefficient defined in Equation~15 of \citet{Padoan:2001} that is a product of polarization efficiency factors, such as the degree of dust alignment and the dust grain polarization cross-section, and~$\psi$ is the angle between the projection of the magnetic field on the plane of the sky and South, and the Stokes parameters are given in the COSMO polarization convention. These equations show the dependence of the dust polarization fraction on~$\gamma$,~$\psi$,~$\alpha$, $\rho$, and~$\epsilon$ and their variations along the line of sight. At large angular scales and away from the Galactic plane, the variation in the observed dust polarization fraction is dominated by~$\gamma$ and~$\psi$ \citep{Hensley:2019, PlanckCollaboration:2020}.

To maximize the chances of detecting this correlation, we limit our analysis to sightlines where the dust in the \lb\ wall contributes the most to the total extinction and where we have high signal-to-noise ratio measurements of the dust polarization fraction~(SNR$_p~>~3$; Figure \ref{fig:snr}). We then select sightlines in the highest quartile of~$A_{V}^{\rm LB}~/~A_{V}^{\rm Planck}$, i.e., sightlines where the extinction in the \lb\ wall has the highest contribution to the total observed extinction (\secref{subsec:red}). For~$A_{V}^{\rm LB}$, we integrate the~3D dust differential extinction maps of \citet{Lallement:2019} in the \lb\ wall. We integrate over~50~pc in each direction, starting with the distance to the inner \lb\ surface as defined by the~3D model of \citet{Pelgrims:2020}. Note that the threshold that corresponds to the highest quartile is~$A_{V}^{\rm LB}~/~A_{V}^{\rm Planck}~\sim~0.2$. This does not necessarily mean that~$\gtrsim~20\%$ of the total extinction is attributable to the \lb\ wall, as the integration over 50~pc may not represent the true Bubble thickness for all sightlines, and~$A_{V}^{\rm Planck}$ is estimated through a scale factor multiplied by the GNILC dust optical depth map (\secref{subsec:red}). Nevertheless, this represents a best estimate of the sky regions for which the \lb\ wall accounts for the largest fraction of the total extinction. This map and the final mask are shown in \figref{fig:masks_LB}. While we define this mask to maximize the chance of detecting an imprint of the \lb\ wall on the dust polarization, we also perform this analysis with different masks. If we redefine $A_{V}^{\rm Planck}$ using the integral from the \lb\ surface to 100~pc, rather than 50~pc, 97\% of the selected sightlines remain identical. The threshold that corresponds to the highest quartile increases to~$A_{V}^{\rm LB}~/~A_{V}^{\rm Planck}~\sim~0.4$ in this case. We additionally perform this analysis with masks based on simple latitude cuts, including one focusing only on sightlines with~$|b|~>~60^\circ$. We find that the conclusions in \secref{subsec:pol} do not change for these different masks.

\subsection{Magnetic Fields in the Local Bubble Wall} \label{subsec:BLB}

\begin{figure}[t!]
\includegraphics[width=\columnwidth]{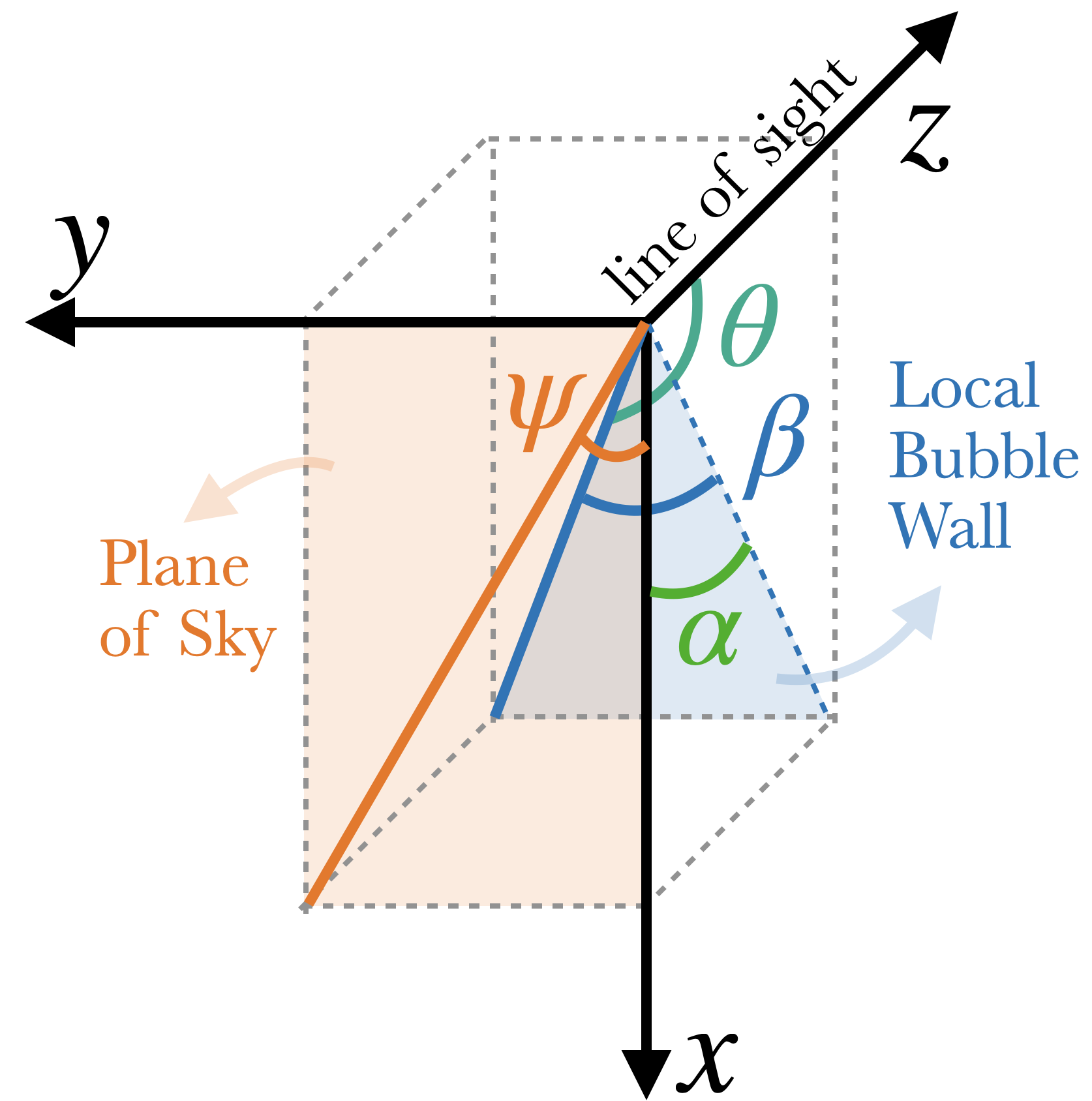}
\caption{Diagram of the angles described and used in \secref{subsec:BLB} for projecting the plane-of-sky magnetic field orientation (orange line) onto the \lb\ wall (blue). The coordinates in this diagram follow the COSMO (HEALPix) convention, which is used in the Planck GNILC maps. These are not the Galactic coordinates used in \figref{fig:slices}. For each position in the sky looking outwards, the local x-axis points South, the local y-axis points East, and the local z-axis points outwards.
\label{fig:angles}}
\end{figure}

We perform the following steps to estimate the angle between the line of sight and the \lb\ wall.

\begin{enumerate}
    \item For the \lb\ geometry, we use the~3D model developed by \citet{Pelgrims:2020}. Since this model is constructed based on the~3D dust map of \citet{Lallement:2019}, we use this dust map in this subsection. Example slices of the map with the model overlaid are shown at the top of \figref{fig:slices}. The resolution of this dust map is~25~pc. We use a~3D Gaussian kernel with a standard deviation of~25~pc to smooth the data as done in \citet{Pelgrims:2020}. This smooths out spurious small-scale fluctuations in the data product that may affect the results.
    \item We calculate the gradient of the differential extinction data cube,~$\nabla A_{{V},i}'$ for each voxel~$i$. This is a vector in the direction of the steepest change in the~3D volume at a given voxel. For the voxels at the surface of the \lb\ wall,~$\nabla A_{{V},i}'$ would therefore be orthogonal to that surface.
    \item For each voxel~$i$, we calculate the angle between the \lb\ surface when projected onto the~$x-z$ plane and the plane of the sky as
    \eq{\alpha_i~=~\arccos{\left(\frac{\nabla A_{{V},i}' \cdot \vec{r_i}}{|\nabla A_{{V},i}'||\vec{r_i}|}\right)}, \, \alpha_i~\in~[0,\pi/2],} 
    where~$\vec{r_i}$ is the line of sight vector, using the Sun, which is at the center of the data cube, as the origin. This angle is shown in \figref{fig:angles} in green.
    \item \citet{Pelgrims:2020} models the distance to the \lb\ surface for each direction on a HEALPix sphere. We sample our~3D Cartesian cube of~$\alpha_i$ at the radial distance defined by the \citet{Pelgrims:2020} model for each line-of-sight direction with a HEALPix pixelization scheme.
    \item Using the Planck GNILC Stokes~$Q$ and~$U$ maps at~353~GHz, we calculate the plane-of-sky magnetic field orientation as 
    \eq{\psi~=~\frac{1}{2} \arctan{\frac{-U}{-Q}}.}
    This angle is shown in \figref{fig:angles} in orange.
    \item We project the plane-of-sky magnetic field orientation onto the \lb\ surface as 
    \eq{\beta~=~\arctan{(\tan{\psi}\cos{\alpha})}.}
    This angle is shown in \figref{fig:angles} in blue.
    \item We calculate the angle between the line of sight and the magnetic field lines tangential to the surface of the \lb\ as 
    \eq{\theta~=~\arccos{(\cos{\beta}\sin{\alpha}).} \label{eq:theta}}
    This angle is shown in \figref{fig:angles} in teal.
\end{enumerate}

\begin{figure}[t!]
\plotone{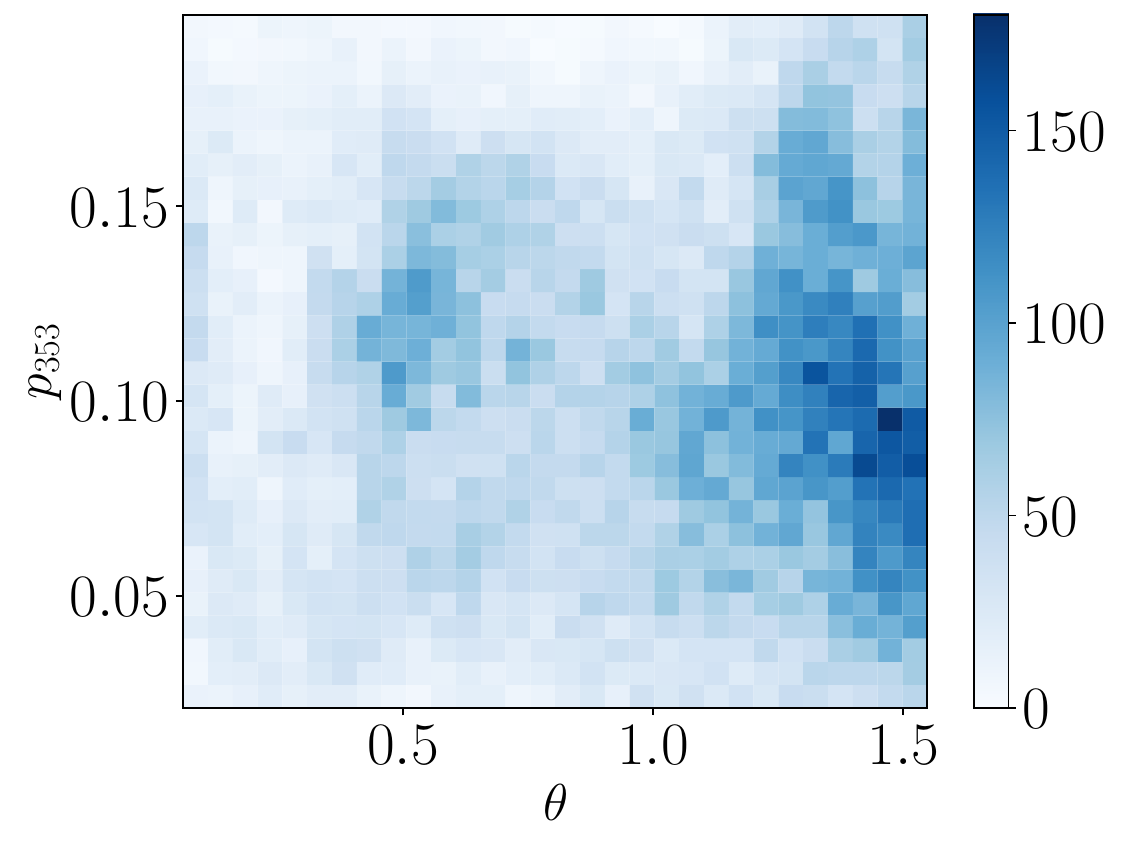}
\caption{A~2D histogram showing the joint distribution of the dust polarization fraction and~$\theta$, the angle between the line of sight and the plane tangent to the surface of the \lb\ (\eqref{eq:theta}) for the sightlines within the mask in \figref{fig:masks_LB}. There is no significant correlation between these two quantities.
\label{fig:ptheta}}
\end{figure}
 
\subsection{No Detected Imprint of the Local Bubble Wall on the Dust Polarization Fraction} \label{subsec:pol}
We do not find any correlation between~$p_{353}$ and~$\theta$ from \eqref{eq:theta} over the mask defined in \figref{fig:masks_LB}. The joint distribution is shown as a~2D histogram in \figref{fig:ptheta}. The Spearman rank coefficient, which is agnostic to the functional dependence between~$p_{353}$ and~$\theta$, is~$3~\times~10^{-3}$. This lack in correlation persists for each of the masks described in \secref{subsec:LBtest}.
This indicates that at least one of the assumptions described in \secref{subsec:assump} is not valid. In other words, either the magnetic field is not generally tangent to the \lb\ surface, the dust polarization is not dominated by the dust within~50~pc of the surface as defined by the \citet{Pelgrims:2020}, or a combination of these possibilities.

We also test whether we find an anti-correlation between~$\theta$ and the absolute value of the Faraday rotation measure (\secref{subsec:rotmeas}) but do not find any evidence for it. Therefore, we do not find an imprint of the detailed~3D geometry of the \lb\ wall on the polarization statistics.

\citet{O'Neill:2024} provide a map of the inclination angle between the \lb\ wall as defined by their model and the plane of the sky. We also test for a correlation between this angle and the dust polarization fraction. We use a similar sightline selection criterion, replacing the~\citet{Lallement:2019} map with the~\citet{Edenhofer:2023} map and the \citet{Pelgrims:2020} model with the~\citet{O'Neill:2024} model. We calculate a Spearman rank coefficient of~0.04, i.e., we do not find any correlation.

\section{Imprint of Dust Complexity on Dust Polarization Fraction} \label{sec:3ddust}
We continue the investigation of how the~3D distribution of dust impacts the dust polarization fraction beyond the \lb, taking into account the distribution of the dust in an extended volume around the Sun. We use the~3D dust maps constructed by \citet{Edenhofer:2023} in this section. The benefits of these maps are that they have high resolution and extend radially up to~1.25~kpc away from the Sun.

If contributions to the polarized dust emission originate from regions along the line of sight with differently oriented magnetic fields, the integrated signal will be depolarized relative to emission from a region with uniform magnetic fields. We postulate that sightlines with multiple dust components that are separated in distance and contribute similarly to the total column density are more likely to have substantial dust emission originating from regions with differently oriented magnetic fields. We test the hypothesis that on average, for sightlines with the same column density, the ones for which the dust is distributed into multiple components at different distances with similar contributions to the total column density are associated with higher levels of depolarization than those for which the dust contribution to the total column density is concentrated. Another way to state our hypothesis is that given two sightlines at the same column density, the one with a more complex~3D dust distribution will have a lower dust polarization fraction on average.

\subsection{Sightline Selection} \label{subsec:select}
We select sightlines with high-fidelity dust polarization measurements (SNR$_p~>~3$), that pass through regions with trustworthy~3D dust reconstruction, and that have a dominant contribution from the dust extinction within the~3D dust maps $A_{V}^{\rm Edenhofer}$ to the total estimated extinction $A_{V}^{\rm Planck}$ (\secref{subsec:red}). The last constraint is to avoid sightlines at the lowest Galactic latitudes, where the dust extends in distance well beyond the regions where the dust is mapped in~3D out to 1.25~kpc. We mask sightlines where the ratio~$A_{V}^{\rm Edenhofer}/A_{V}^{\rm Planck} < 0.5$.

\begin{figure}[t!]
\plotone{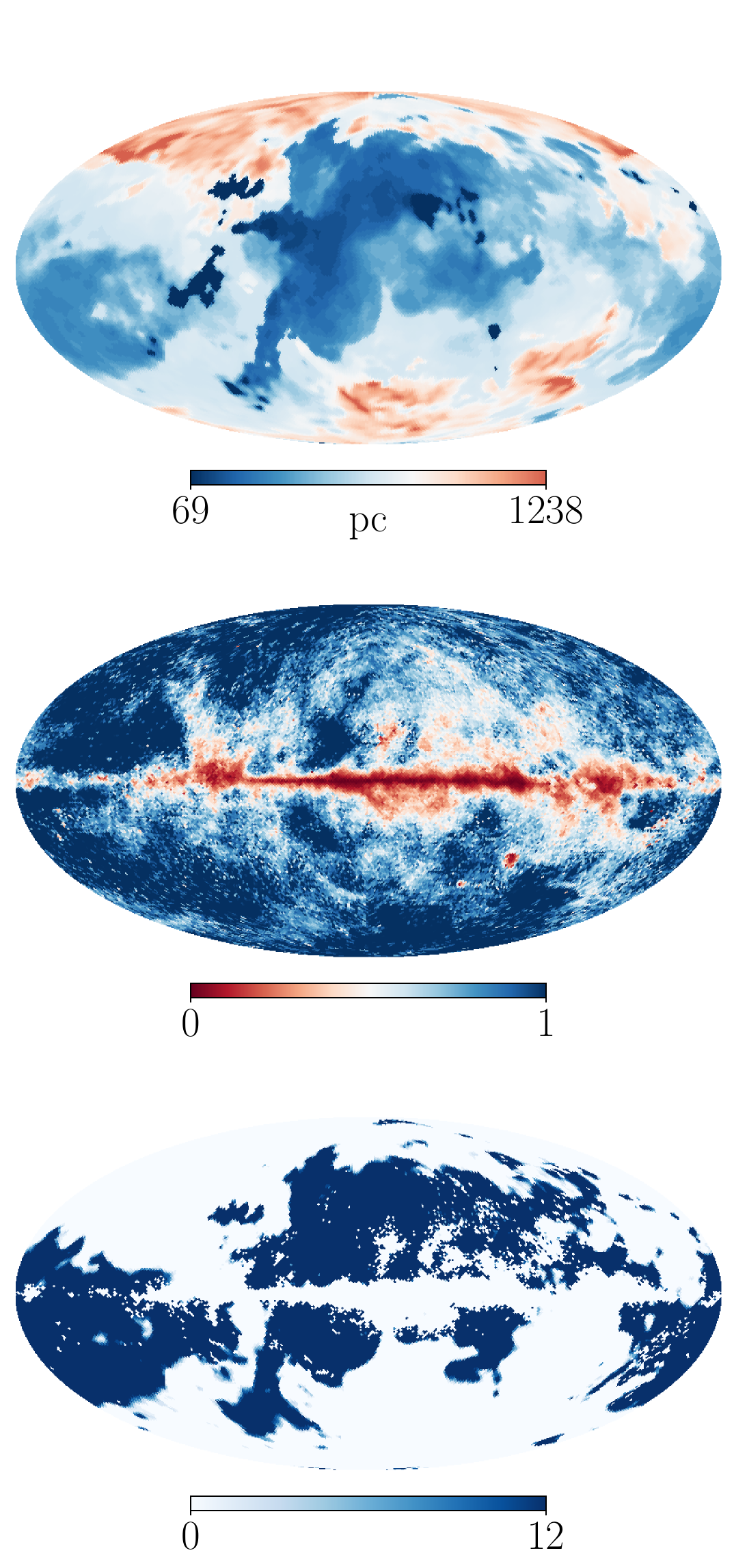}
\caption{Maps used for selecting sightlines for the analysis described in \secref{sec:3ddust}. \textit{Top panel:} A map of the distance at which the extinction in the first posterior sample of the \citet{Edenhofer:2023}~3D dust maps reaches~50~mmag. This is plotted with a diverging colorbar centered on~200~pc, the cutoff we use in our sightline selection, with the allowed regions shown in blue. \textit{Middle panel:} A map of the ratio of the \citet{Edenhofer:2023}~3D dust differential extinction map integrated out to~1.25~kpc over the Planck dust extinction. This is plotted with a diverging colorbar centered on 0.5, the cutoff we use in our sightline selection, with the allowed regions shown in blue. \textit{Bottom panel:} A sum of the masks of the selected regions over each of the~12~posterior samples, combining the selected regions from the quantities in the top panel, middle panel, and \figref{fig:snr}.
\label{fig:masks_complexity}}
\end{figure}

\citet{Edenhofer:2023} use estimates of stellar distances and extinctions from \citet{Zhang:2023} to construct their~3D dust maps. However, \citet{Edenhofer:2023} found that their estimated extinction disagrees with \citet{Zhang:2023} where the radially integrated differential extinction is below~50~mmag or above~4~mag. We therefore use those thresholds to mask sightlines where the dust differential extinction is likely to be significantly over- or under-estimated. The distance to those thresholds for a given sightline varies slightly for the~12~different posterior samples, so we apply a slightly different mask to each posterior sample. We only consider sightlines where the differential extinction integrated radially outwards reaches~50~mmag within~200~pc of the Sun. A threshold higher than~200~pc would include more sightlines in the selection, but it would shorten the minimum path length considered. Because in \secref{subsec:res}, we compare our results with the \citet{Leike:2020} maps, which only extend out to~270~pc, we find~200~pc to be a good balance between having a large enough sample size and minimum path length. The minimum path length considered through the \citet{Edenhofer:2023} maps for estimating the dust complexity is therefore 1.05~kpc.

A map of the distance at which the extinction reaches~50~mmag for different sightlines is shown at the top of \figref{fig:masks_complexity}. This is for the first posterior sample, but the equivalent maps for the remaining~11~posterior samples look visually indistinguishable. We center the diverging colorbar in this subplot to~200~pc to show which sightlines pass the threshold.

After masking sightlines with~SNR$_p~<~3$ or with~$A_{V}^{\rm Edenhofer}/A_{V}^{\rm Planck} < 0.5$ for the~12 posterior samples, the integrated differential extinction within~1.25~kpc is higher than~4~mag for only about~10 sightlines per posterior sample. We exclude these sightlines from our analysis.

After performing all the cuts, we are left with about~19,100 sightlines per posterior sample ($\sim40\%$ of the sky) for this analysis. Those sightlines are shown at the bottom of \figref{fig:masks_complexity}. The map shown is the sum of the binary masks over the~12~posterior samples. Note that the only discrepancies between the~12 posterior samples are at the edges of the selected regions. The rest of the map is either~12 or~0.

\subsection{Dust Complexity} \label{subsec:comp}
We aim to quantify the complexity of the~3D dust distribution along each line of sight. For this, we take inspiration from \citet{Panopoulou:2020}, who perform a Gaussian decomposition of the neutral hydrogen (\hi) emission spectra of each sightline to quantify its complexity. They use these components~$i$, weighted by their column density~$N^i_{\rm HI}$, to define a metric
\eq{\mathcal{N}_c^{\rm HI}~=~\sum_{i=1}^{{\rm n}_{\rm clouds}}\frac{N^i_{\rm HI}}{N^{\rm max}_{\rm HI}}, \label{eq:nchi}}
where~$N^{\rm max}_{\rm HI}$ is the column density of the component with highest~$N_{\rm HI}$. This is a more relevant measurement of complexity than simply counting the number of components because it takes into account the relative contribution of each detected component, and the dust polarization signal arises from a density-weighted integral along the line of sight. If there are two components along the line of sight and they have equal column densities, then~$\mathcal{N}_c^{\rm HI}~=~2$, whereas if one has half the column density of the other, then~$\mathcal{N}_c^{\rm HI}~=~1.5$, and so forth. Therefore,~$\mathcal{N}_c^{\rm HI}$ could be, for instance,~2 for any number of components larger than~1. This metric was used to detect an imprint of line-of-sight magnetic field tangling in dust polarization in \citet{Pelgrims:2021}.

We use a metric inspired by \eqref{eq:nchi} to quantify the complexity of the~3D dust distribution, i.e., we decompose the~3D dust sightlines into ``clouds," or components along the line of sight, and then use a version of \eqref{eq:nchi}, replacing~$N_{\rm HI}$ with~$N_{\rm H}$ inferred from the dust extinction (\secref{sec:data}), i.e.,
\eq{\mathcal{N}_c~=~\sum_{i=1}^{{\rm n}_{\rm clouds}}\frac{N^i_{\rm H}}{N^{\rm max}_{\rm H}}. \label{eq:nc}}

To decompose each of the~19,100 sightlines into different components, we use the dendrogram technique \citep{Rosolowsky:2008}. We use the Python package \texttt{astrodendro}\footnote{\url{http://www.dendrograms.org/}}. \citet{Cahlon:2023} applied the~3D version of this technique on the \citet{Leike:2020}~3D dust maps to produce a uniform catalog of molecular clouds in the Solar neighborhood. In this analysis, we apply the~1D version on individual sightlines.

Dendrogramming identifies density peaks in the data and connects them along hierarchical isosurfaces of constant~$n_{\rm H}$, forming a tree-like structure. We refer the reader to \citet{Rosolowsky:2008} for a description of the core algorithm. We focus the explanation here on the algorithm's three parameters. The first parameter defines the minimum absolute Hydrogen number density~$n_{\rm H}$ threshold for a structure to be included as part of the tree. We set this parameter to~0. This is because the tree-like structure constructed by this algorithm is not useful for our purposes. We only consider the density peaks identified.

The remaining two parameters,~$\Delta_n$ and~$\#_{\rm voxels}$, define the minimum prominence for a peak to be considered an independent component. Its largest~$n_{\rm H}$ has to be~$\Delta_n$ above the~$n_{\rm H}$ of the adjacent isosurface for it to be considered an independent component from that isosurface. Similarly,~$\#_{\rm voxels}$ defines the threshold number of voxels it has to span along a sightline to be considered an independent component, where each voxel spans~7~pc of the sightline. If a peak passes both of those thresholds, it is identified as a component by the algorithm. 

For this analysis, we experiment with various values for~$\Delta_n$ and~$\#_{\rm voxels}$ and plot the resulting peaks identified for each variation. We find a range between~$\Delta_n~=~1.94~\times~10^{-3}~{\rm cm}^{-3}$ ($A_{V}'~=~3~\times~10^{-6}$~mag/pc) and~$\Delta_n~=~3.87~\times~10^{-3}~{\rm cm}^{-3}$ ($A_{V}'~=~9~\times~10^{-6}$~mag/pc) and between~$\#_{\rm voxels}~=~3$ (21~pc) and~$\#_{\rm voxels}~=~5$ (35~pc) recovers the visually identified peaks. We report our results for the fiducial values of~$\Delta_n~=~4.52~\times~10^{-3}~{\rm cm}^{-3}$ ($A_{V}'~=~7~\times~10^{-6}$), and~$\#_{\rm voxels}~=~3$ (21~pc). However, we find our results to be robust to all the variations we test for within the ranges mentioned.

\begin{figure}[t!]
\includegraphics[width=\columnwidth]{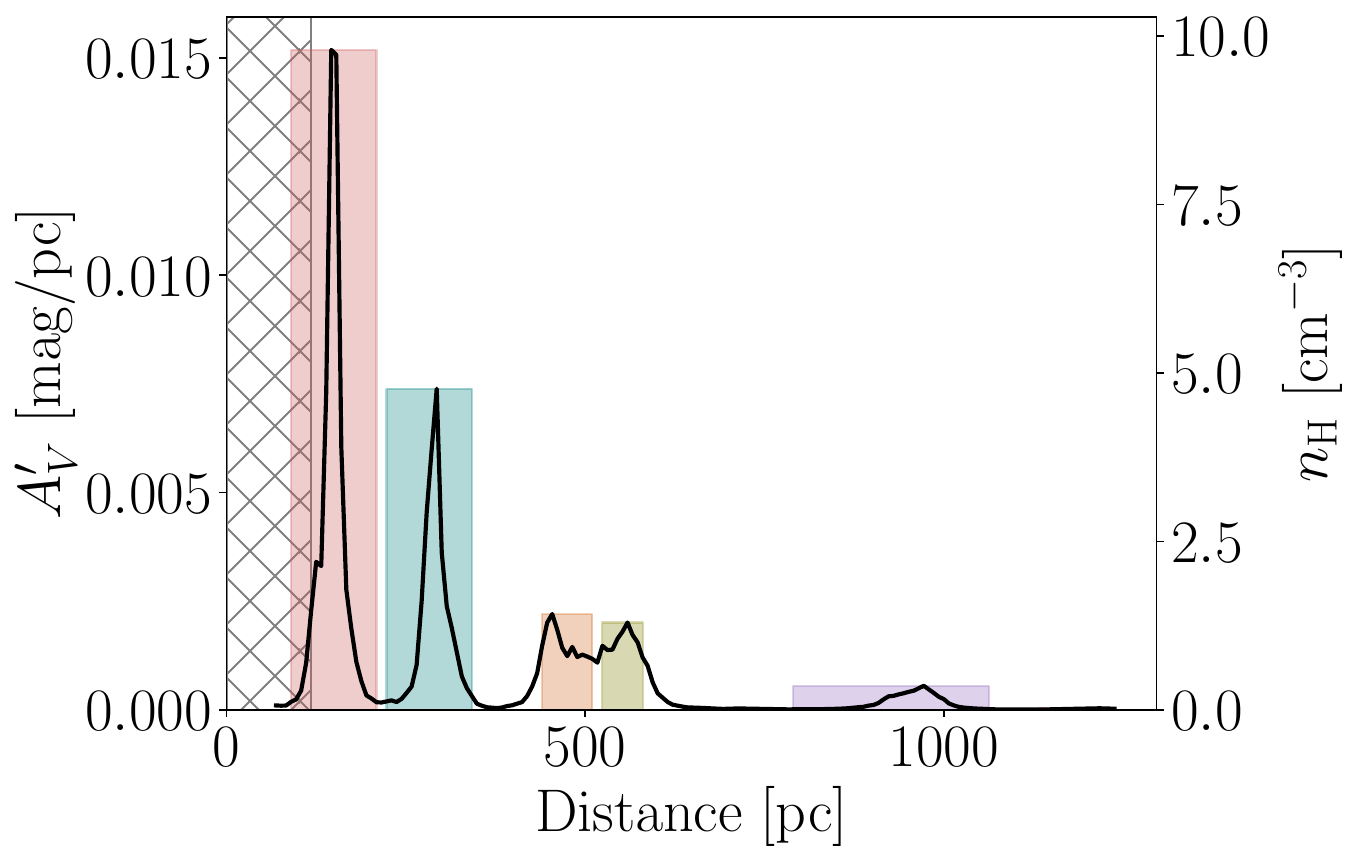}
\caption{The dust distribution in units of differential extinction (mag/pc, left y-axis) and equivalent Hydrogen number density (cm$^{-3}$, right y-axis) along a representative sightline through the \citet{Edenhofer:2023} maps. The Galactic coordinates of this sightline are~$l~=~163.12^\circ$ and~$b~=~-11.42^\circ$. The region before the extinction reaches~50~mmag (hatched) is discarded from our analysis. The components identified by the dendrogram algorithm with~$\Delta_n~=~4.52~\times~10^{-3}~{\rm cm}^{-3}$ ($A_{V}'~=~7~\times~10^{-6}$), and~$\#_{\rm voxels}~=~3$ (21~pc) are shaded in different colors.
\label{fig:sightline}}
\end{figure}

We run the dendrogram algorithm on each sightline separately, considering the entire sightline from~69~pc to~1.25~kpc. However, we only keep identified components with peaks that are radially farther than the distance at which the extinction in that sightline reaches~50~mmag (\secref{subsec:select}). If the distance of the peak of a component is farther than this threshold but part of the component is below that threshold (\figref{fig:sightline}), we still consider the part of the component that is below that threshold when integrating over the~$n_{\rm H}$ of that component. We find that the peak with the highest dust column density lies within~270~pc from the Sun for most sightlines within our mask.

\begin{figure}[t!]
\includegraphics[width=\columnwidth]{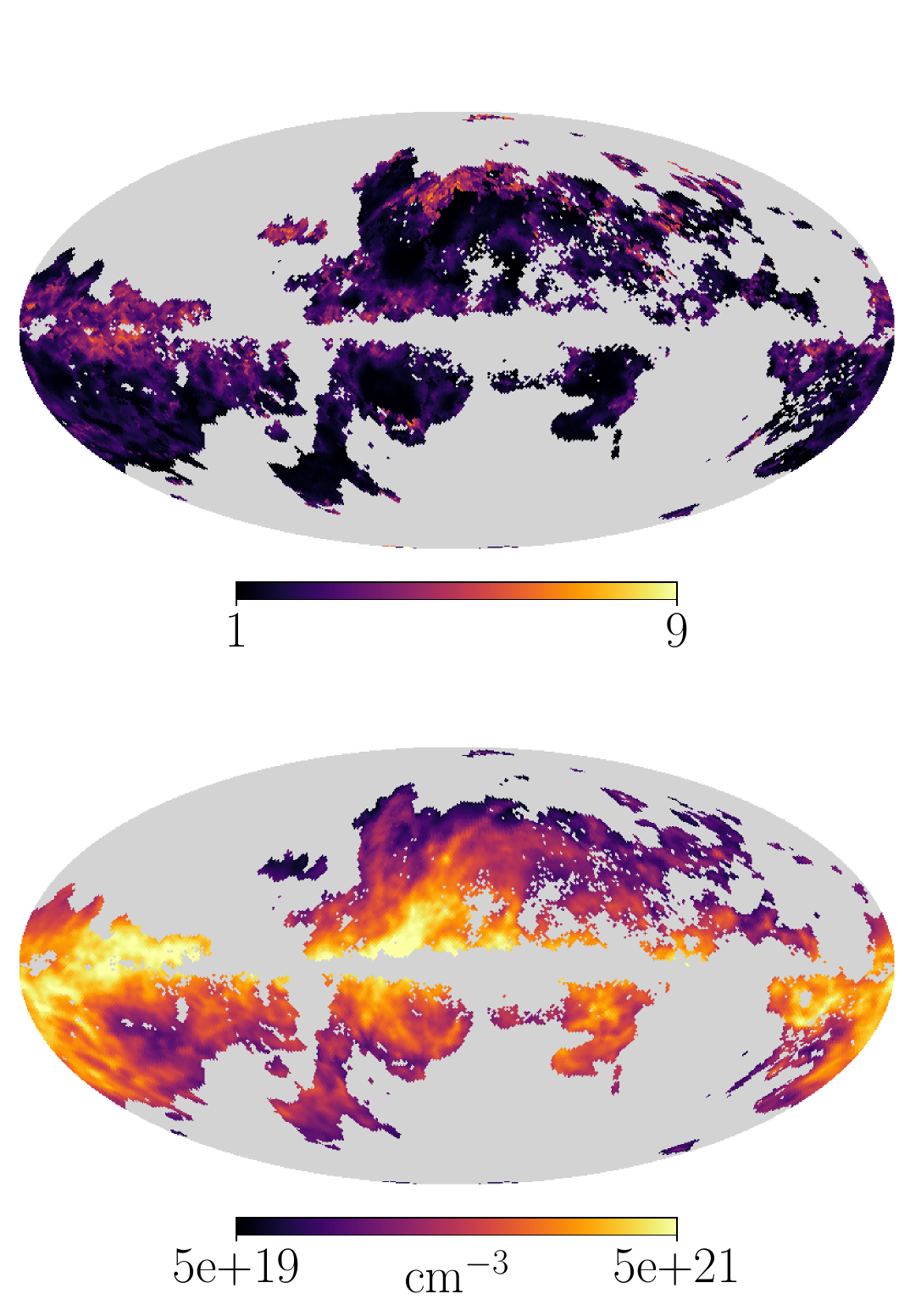}
\caption{\textit{Top panel:} A map of~$\mathcal{N}_c$ calculated using the dendrogram algorithm with~$\Delta_n~=~4.52~\times~10^{-3}~{\rm cm}^{-3}$ ($A_{V}'~=~7~\times~10^{-6}$), and~$\#_{\rm voxels}~=~3$ (21~pc). \textit{Bottom panel:} A log-scale map of~$N_{\rm H}^{\rm Edenhofer}$ formed by integrating over the entire dataset (up to~1.25~kpc) with units of~${\rm cm}^{-2}$. These maps are shown only for the sightlines selected in \secref{subsec:select} and for one of the posterior samples of \citet{Edenhofer:2023}.
\label{fig:nc_nh_maps}}
\end{figure}

We calculate \eqref{eq:nc} for each sightline to quantify its dust complexity. For the sightline in \figref{fig:sightline}, for example,~$\mathcal{N}_c$=2.16. We show a map of~$\mathcal{N}_c$ for the sightlines we select in \secref{subsec:select} for one of the~12 posterior samples of the \citet{Edenhofer:2023}
maps at the top of \figref{fig:nc_nh_maps}.\footnote{We make the~$\mathcal{N}_c$ maps for the~12 posterior samples of \citet{Edenhofer:2023} publicly available at \url{https://doi.org/10.7910/DVN/IW09AE} \citep{DVN/IW09AE_2024}}

\subsection{Nearest-Neighbor Matching} \label{subsec:nnm}
To examine the effect of line-of-sight dust complexity on the dust polarization fraction, we compare sightlines that have the same total column densities but very different~3D dust distributions. We define low- and high-dust complexity sightlines as sightlines with~$\mathcal{N}_c\leq1.1$ and~$\mathcal{N}_c\geq1.5$, respectively. The goal is to compare sightlines with different dust distributions, so the particular~$\mathcal{N}_c$ threshold values are less important. We start with these values for our fiducial analysis because they are similar to the thresholds used in \citet{Pelgrims:2021}. The \citet{Pelgrims:2021} analysis used a Gaussian decomposition of~3D maps of the neutral hydrogen emission line, where the third dimension is radial velocity. They used~$\mathcal{N}_c$=1 and~$\mathcal{N}_c\geq1.5$ in their analysis. However, for the data and mask we use in our analysis, we find that across the 12 posterior sample maps from \citet{Edenhofer:2023}, only 8-31 sightlines have~$\mathcal{N}_c$=1, i.e., a single dust component. Since this does not represent a large enough sample size, we define the low-complexity bin as sightlines with~$\mathcal{N}_c\leq1.1$. This increases the number of sightlines in that bin to around~1,000 per posterior sample. The number of sightlines in the~$\mathcal{N}_c\geq1.5$ bin is around~13,000. However, we also confirm that our results are robust to variations in these thresholds (\secref{subsec:valid}).

Because the~$\mathcal{N}_c$ bins are widely separated, small fluctuations in~$\mathcal{N}_c$ due to the choices of~$\Delta_n$ and~$\#_{\rm voxels}$ do not cause sightlines from one bin to shift to the other bin. However, we also verify that our results are robust to different choices of~$\Delta_n$ and~$\#_{\rm voxels}$.

We treat the column density as a confounding variable when comparing the distribution of~$p_{353}$ for the low- and high-complexity bins. Any difference in the~$p_{353}$ distributions for the low- and high-complexity bins could potentially be explained by a difference in the column density integrated over the distance used to calculate~$\mathcal{N}_c$ or a difference in the total column density over a sightline. Therefore, both need to be taken into account for a fair comparison of~$p_{353}$ between the the two complexity bins.

We integrate~$n_{\rm H}$ for each sightline up to~1.25~kpc, starting from either the distance at which the extinction reaches~50~mmag or the minimum distance of the first detected component whose peak lies farther than that distance as explained in \secref{subsec:comp}, whichever is closer. We call this~$N_{\rm H}^{\rm Edenhofer}$ and show a map of it for the sightlines considered in our analysis in the bottom panel of \figref{fig:nc_nh_maps}.

Since~$N_{\rm H}^{\rm Edenhofer}$ only takes into account dust up to~1.25~kpc, we also consider the total column density over a sightline. We convert~$A_{V}^{\rm Planck}$ (\secref{subsec:red}) to~$N_{\rm H}^{\rm Planck}$ following the formalism in \secref{subsec:dustdata}. In \secref{subsec:valid}, we also experiment with adding the absolute value of the Galactic latitude as an additional confounding variable.

We perform nearest-neighbor matching with no replacement between the sightlines in the low- and high-complexity bins. We pair up each low-complexity sightline to the high-complexity sightline with the closest~$N_{\rm H}^{\rm Edenhofer}$ and~$N_{\rm H}^{\rm Planck}$ values based on the Manhattan distance using the ball-tree algorithm. The number of pairs is, therefore, equal to the number of sightlines in the low-complexity group, which is the smaller group. For each matched pair, we subtract the~$p_{353}$ of the sightline with the lower complexity from the~$p_{353}$ of the sightline with the higher complexity. We take the average of the differences over the matched pairs,~$\Delta p_{353}$, to test whether sightlines with higher dust complexity have higher depolarization levels on average than sightlines with lower dust complexity, i.e.,~$\Delta p_{353} < 0$. 

\subsection{Statistical Tests} \label{subsec:stats}
We determine the statistical significance of our results through permutation tests. We perform the analysis described in \secref{subsec:nnm} and obtain~$\Delta p_{353}^s$ as the mean over the pairs for each of the~12 posterior samples~$s$. We perform a permutation-based null test in which we randomly choose one slightline to subtract from the other in each pair, rather than always subtracting~$p_{353}$ of the low-complexity sightline. We repeat this~10,000 times and obtain a distribution of~$\Delta p_{353}^{s,\,{\rm null}}$ for each posterior sample~$s$. We calculate a p-value as the proportion of~$\Delta p_{353}^{s,\,{\rm null}}$ that are equal to or more extreme than~$\Delta p_{353}^s$.

We additionally use an alternative null test. Instead of separating sightlines into high- and low-complexity groups, we randomly select~25\% of sightlines (5,050 sightlines) to be in one group and~25\% to be in the other group. We then run the same analysis on those~2 groups, pairing them up based on~$N_{\rm H}^{\rm Edenhofer}$ and~$N_{\rm H}^{\rm Planck}$ and subtracting~$p_{353}$ of one group from that of the other for each pair. To ensure we are not biasing the null test by matching neighboring sightlines which may have similar~$p_{353}$, for each run, we randomly alternate between selecting the sightlines in one group to be in the Northern Galactic hemisphere and the sightlines in the other group to be in the Southern Galactic hemisphere and the other way around. We find that sightlines in the Northern Galactic hemisphere tend to have higher~$p_{353}$ on average than sightlines in the Southern Galactic hemisphere, so~$\Delta p_{353}^{s,\,{\rm null}}$ will be biased towards positive or negative values based on which of the two groups is selected from which hemisphere. Using this version of the null test, therefore, takes this bias into account and yields a more conservative estimate of the significance of our hypothesis test. The permutation-based null test, however, guarantees the same number of pairs as the hypothesis test, so we use that version as the main null test when reporting the results. We find that our results are consistent regardless of which null test we use.

We consider results with a two-tailed p-value~$<0.001$ to be statistically significant. We repeat this analysis for several reasonable values for~$\Delta_n$ and~$\#_{\rm voxels}$ and for different thresholds of the~$\mathcal{N}_c$ bins to ensure that our results are independent of those choices.

In addition to checking the statistical significance of~$\Delta p_{353}^s$ for each of the~12 posterior samples, we also report the mean and standard deviation of~$\Delta p_{353}^s$ across the posterior samples. The error propagated due to the uncertainty on~$p_{353}$ is~2 orders of magnitude less than the result. The uncertainty is dominated by the scatter from the~12 posterior samples of the~3D dust map.

\subsection{Results} \label{subsec:res}

\begin{figure}[t!]
\includegraphics[width=\columnwidth]{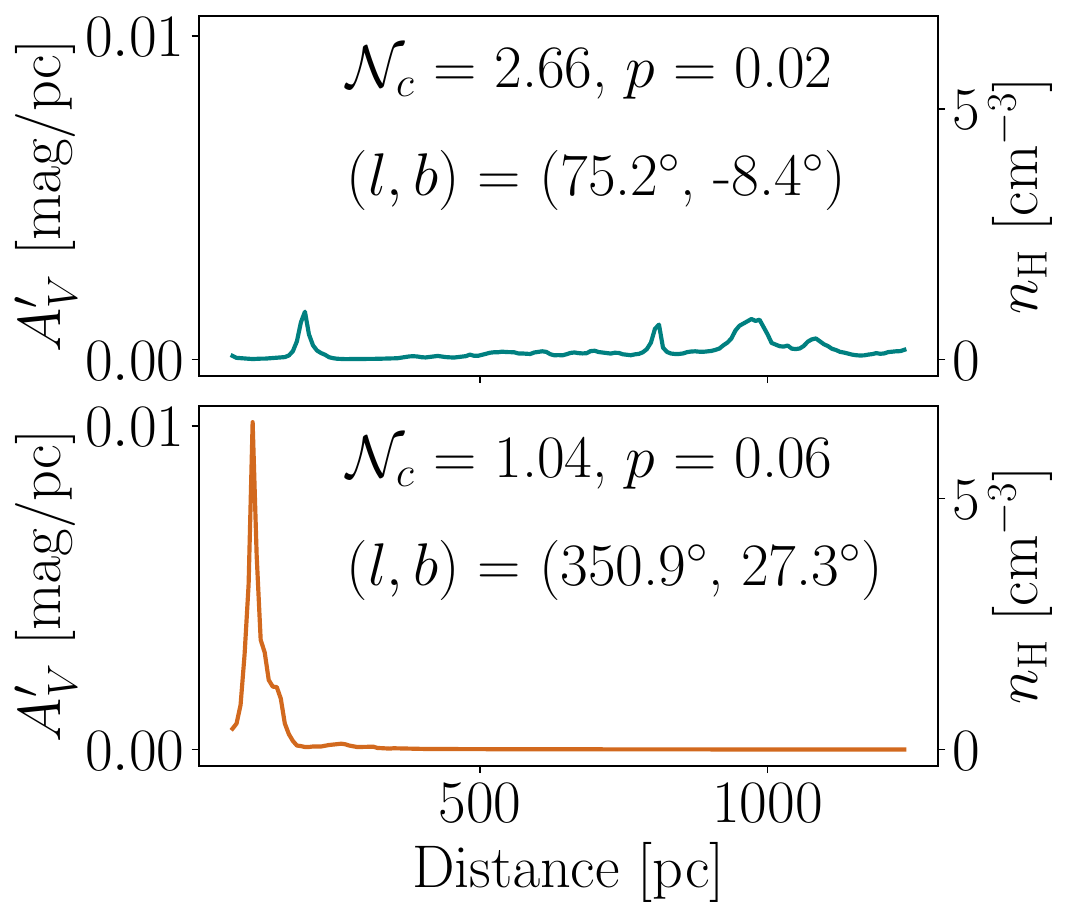}
\caption{The dust distribution in units of differential extinction (mag/pc, left y-axis) and equivalent Hydrogen number density (cm$^{-3}$, right y-axis) along a pair of matched sightlines through the first posterior sample of the \citet{Edenhofer:2023} map. These sightlines have the same~$N_{\rm H}^{\rm Edenhofer}$, but the top one has a higher complexity than the bottom one. The Galactic coordinates, dust complexity, and dust polarization fraction of each of the sightlines are denoted on their subpanels.
\label{fig:pair}}
\end{figure}

\begin{figure}[t!]
\includegraphics[width=\columnwidth]{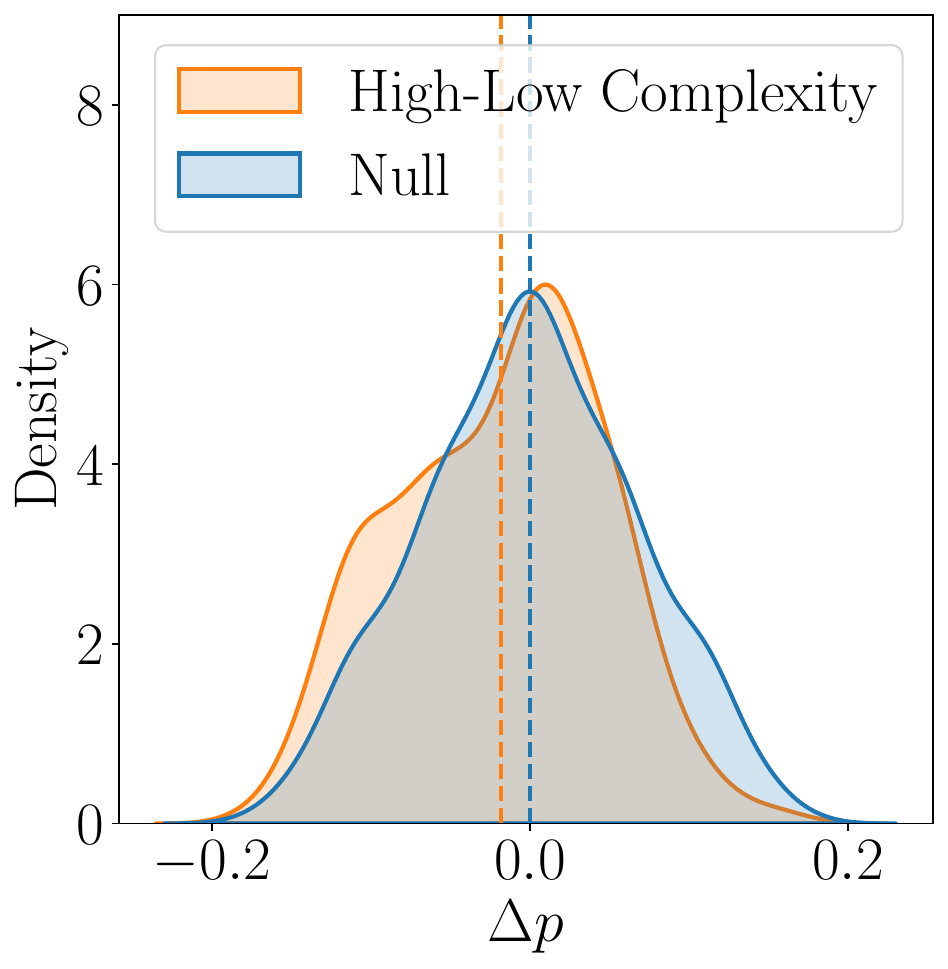}
\caption{Kernel density estimate plots of the~$\Delta p_{353}$ distributions over all matched pairs of sightlines of one posterior sample of the \citet{Edenhofer:2023} maps. The~$p_{353}$ of the lower complexity sightline is always subtracted from that of the higher complexity one in the orange distribution. This has a mean of~$\Delta p_{353} = -1.81~\times~10^{-2}$, which is plotted as an orange dot in \figref{fig:res_all}. The blue distribution contains the same pairs as the orange one with the sign randomly flipped for each pair, i.e., the distribution of the permutation-based null test described in \secref{subsec:stats}.
\label{fig:dp_dist}}
\end{figure}

We perform the analysis described in the previous subsections on the \citet{Edenhofer:2023} maps which extend radially to~1.25~kpc and compare the dust polarization fractions of sightlines with~$\mathcal{N}_c\leq1.1$ and those with~$\mathcal{N}_c\geq1.5$. An example of a pair of matched sightlines with the same~$N_{\rm H}^{\rm Edenhofer}$ and~$N_{\rm H}^{\rm Planck}$ is shown in \figref{fig:pair}. Even though the dust polarization fraction of the higher-complexity sightline is lower than that of the lower-complexity sightline in this example, not all sightlines follow this trend. We are only looking for a statistically significant average effect. An example of the distribution of the differences in~$p_{353}$ over all pairs in one of the map posterior samples is shown in \figref{fig:dp_dist} along with the same distribution for the permutation-based null test.

\begin{figure}[t!]
\includegraphics[width=\columnwidth]{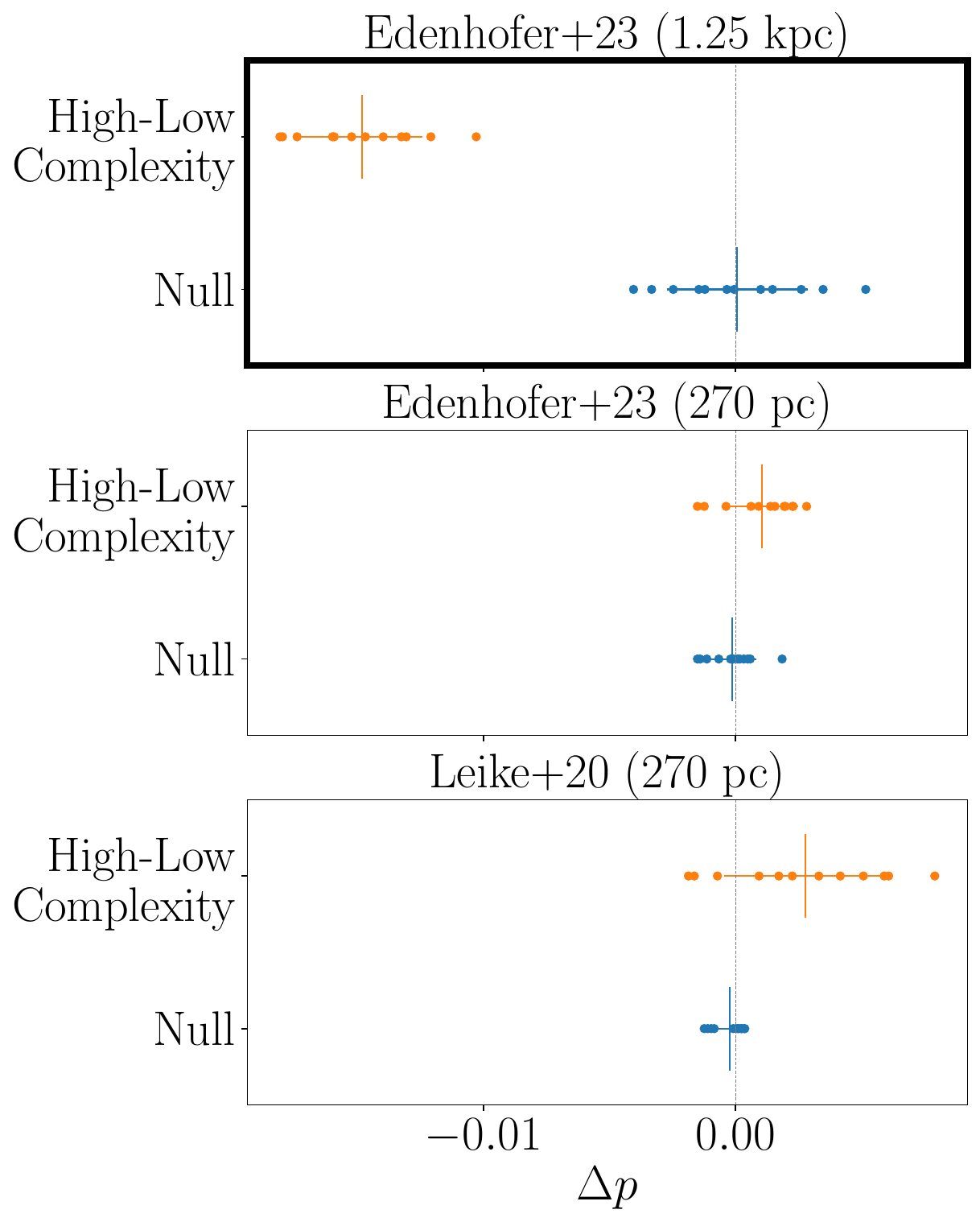}
\caption{The mean difference in~$p_{353}$ over the paired sightlines for each of the~12 posterior samples, where for each pair,~$p_{353}$ for the lower-complexity sightline is subtracted from~$p_{353}$ for the higher-complexity sightline.~$\mathcal{N}_c$ is calculated on the \citet{Edenhofer:2023} maps up to~1.25~kpc (top), the \citet{Edenhofer:2023} maps up to~270~pc (middle), and the \citet{Leike:2020} maps up to~270~pc (bottom). For each panel, the mean differences for the actual test are plotted in orange, and samples from the permutation-based null tests are plotted in blue. The mean and standard deviation of the~12 means for each test are also plotted. The top panel agrees with our hypothesis: that sightlines with similar column densities will, on average, exhibit lower dust polarization fractions when their~3D dust distribution is more complex. 
\label{fig:res_all}}
\end{figure}

For this test,~$\Delta p_{353}^s$, the mean over the paired sightlines for each posterior sample~$s$, is plotted in orange at the top of \figref{fig:res_all}. The mean and standard deviation of~$\Delta p_{353}^s$ over the~12 samples are~$-1.47~\times~10^{-2}$ and~$0.22~\times~10^{-2}$, respectively. All~12 posterior samples pass both null tests described in \secref{subsec:stats} with a p-value~$<0.001$. A random~$\Delta p_{353}^{s,\,{\rm null}}$ for each posterior sample is shown in blue at the top of \figref{fig:res_all} as well. Therefore, we find that higher dust complexity at equivalent column densities is associated with depolarization at the~1.5\% level. This is at the level of~6.8\% of the maximum dust polarization fraction measured by \citet{PlanckCollaboration:2020} at~353~GHz and~80$'$.

To determine whether this result is uniquely enabled by the \citet{Edenhofer:2023} dust maps since they extend radially to~1.25~kpc, we repeat the analysis using the~3D \citet{Leike:2020} dust maps, which extend to~370~pc in the positive and negative Galactic-X and~Y coordinates and~270~pc in the positive and negative Galactic-Z coordinate. For consistency between different sightlines in those maps, we truncate all sightlines at~270~pc. Also, we start each sightline at~70~pc since \citet{Leike:2020} find that the reconstructed dust density closer than~70~pc resembles a smeared-out version of the farther dust, an artifact related to systematic data biases. \citet{Leike:2020} also provide~12 posterior samples for their maps which we use for this analysis. Because we sample these data in increments of~2~pc as opposed to~7~pc as in the case of the \citet{Edenhofer:2023} data, we set the dendrogram parameter~$\#_{\rm voxels}~=~10$, which corresponds to~20~pc, compared to~$\#_{\rm voxels}~=~3$, which corresponds to~21~pc in the case of the \citet{Edenhofer:2023} data. We keep the dendrogram parameter~$\Delta_n$ the same for both maps. This results in about~3,500 sightlines with~$\mathcal{N}_c\leq1.1$ and about~8,500 with~$\mathcal{N}_c\geq1.5$.

The results for repeating the analysis using the \citet{Leike:2020} maps instead are shown at the bottom of \figref{fig:res_all}. The mean and standard deviation of~$\Delta p_{353}^s$ over the~12 posterior samples are~$2.75~\times~10^{-3}$ and~$3.06\times~10^{-3}$, respectively. None of the~12 samples pass either of the null tests described in \secref{subsec:stats}, i.e., they are indistinguishable from the distributions of~$\Delta p_{353}^{s,\,{\rm null}}$. We also show a random~$\Delta p_{353}^{s,\,{\rm null}}$ for each posterior sample in blue in the same subplot of \figref{fig:res_all}.

To determine whether the null result is attributed to using a different dataset or to the lower extent in radial distance, we run the analysis on the \citet{Edenhofer:2023} maps up to~270~pc, the same distance used for the \citet{Leike:2020} maps. The number of sightlines with~$\mathcal{N}_c\leq1.1$ and~$\mathcal{N}_c\geq1.5$ after this distance cut are about~10,000 and~3,500, respectively. The difference in the~$\mathcal{N}_c$ distributions between this data and the \citet{Leike:2020} maps is due to several differences in the maps and the post-processing we perform on them, including having to use slightly different dendrogram parameters and not counting components whose peak is closer than the distance where the differential extinction integrated radially outwards reaches~50~mmag in the \citet{Edenhofer:2023} maps among other differences.
We plot the results in the middle of \figref{fig:res_all}. The mean and standard deviation of~$\Delta p_{353}^s$ over the~12 posterior samples are~$1.05~\times~10^{-3}$ and~$1.36~\times~10^{-3}$, respectively. Again, none of the~12 samples pass either of the null tests described in \secref{subsec:stats}, i.e., they are indistinguishable from the distributions of~$\Delta p_{353}^{s,\,{\rm null}}$. The consistency of this result with null as well indicates that the null result we found using the \citet{Leike:2020} maps is due to only considering distances up to~270~pc, not the choice of~3D dust dataset. This illustrates the role that dust components farther than~270~pc play in affecting polarization measurements. 

\subsection{Validation} \label{subsec:valid}
In this subsection, we summarize some of the tests we performed to verify our results. As described in \secref{subsec:comp}, we find that our results are robust to reasonable variations in~$\Delta_n$ and~$\#_{\rm voxels}$. These are summarized in \tabref{tab:params} in Appendix~\ref{sec:app}. The results are also consistent when using the Planck~R3.01~353~GHz maps instead of the GNILC maps.

To ensure that the pairs were matched correctly, we examine the distributions of the differences in~$N_{\rm H}^{\rm Edenhofer}$ and~$N_{\rm H}^{\rm Planck}$ between the matched pairs. We verify that those differences peak near zero and are not skewed toward the positive or negative values. We find that to be the case for both variables and over all~12~posterior samples of \citet{Edenhofer:2023}. We also find our results to be robust when including~$|b|$ as an additional confounding variable to~$N_{\rm H}^{\rm Edenhofer}$ and~$N_{\rm H}^{\rm Planck}$ in the pair matching -- in other words, the result is not attributable to a dependence of the dust polarization fraction on Galactic latitude. We show these results in \tabref{tab:match} in Appendix~\ref{sec:app}.

We also examine the angular separations between the sightline pairs and the discrepancies in their overall path lengths. The distributions of angular distances are consistent with a random distribution of angular distances for all~12~posterior samples of \citet{Edenhofer:2023}. We also find no significant differences in the total path lengths between the paired sightlines across all 12 samples.

We experiment with varying the~$\mathcal{N}_c$ thresholds that define the bin edges of the low- and high-complexity sightlines. The~25th and~75th percentiles of the~$\mathcal{N}_c$ distribution over our mask vary slightly over the~12~posterior samples of \citet{Edenhofer:2023} and over variations in~$\Delta_n$ and~$\#_{\rm voxels}$. However, they are roughly~1.4 and~2.3, respectively. Therefore, we split our sightlines based on these values into low- ($\mathcal{N}_c\leq1.4$) and high-complexity ($\mathcal{N}_c\geq2.3$) bins to achieve a roughly equal number of sightlines in each group. This should improve the pair-matching outcomes since there are more sightlines to match from in the smaller group. We find that our results are robust to this change as shown in \tabref{tab:params} in Appendix~\ref{sec:app}.

Since neighboring sightlines are likely to have similar~$\mathcal{N}_c$ and similar~$p_{353}$ values, we test whether large regions of neighboring sightlines belonging to either the low- or high-complexity groups bias our results. We randomly sample~1,000 sightlines out of about~5,000 sightlines from each of the low- and high-complexity groups before pair matching, where we use~$\mathcal{N}_c\leq1.4$ and~$\mathcal{N}_c\geq2.3$ for these groups, respectively, in this case. We find that our results are robust to this test as shown in \tabref{tab:samp} in Appendix~\ref{sec:app}.

Finally, we modify the definition of~$\mathcal{N}_c$ from \eqref{eq:nc} to
\eq{\mathcal{N}_c = \frac{N_{\rm H}^{\rm Edenhofer}}{N^{\rm max}_{\rm H}},}
i.e., we use the total column density out to~1.25~kpc rather than a sum over the dust components in the numerator. With the new definition,~$\mathcal{N}_c$ is only sensitive to the dendrogram-identified component with the highest column density~$N^{\rm max}_{\rm H}$ rather than also being sensitive to the other dendrogram-identified components. Since the dendrogram parameters define the minimum prominence for a peak to be considered an independent component, we expect~$N^{\rm max}_{\rm H}$ to be the least sensitive component to those parameters. Therefore, the new definition is much less sensitive to the dendrogram parameters. The~25th and~75th percentiles of this modified version of~$\mathcal{N}_c$ are about~0.05 and~0.09 for the~12 posterior samples, and we use these as the upper and lower thresholds for the low- and high-complexity bins, respectively. Over the~12 posterior samples, we find a mean and standard deviation of~$\Delta p_{353}~=~0.96~\times~10^{-2}~\pm~0.34~\times~10^{-2}$. This passes the null test with a p-value$~<~0.001$.

\section{Discussion and Conclusions} \label{sec:conc}
In this paper, we explore how different geometrical factors affect the fractional polarization of the dust emission. In \secref{sec:LB}, we test whether we detect an imprint of the \lb\ geometry on the dust polarization fraction. Following the well-motivated assumptions that the magnetic field lines are tangential to the \lb\ surface and that this surface is defined by the model of \citet{Pelgrims:2020}, we test for a correlation between the measured dust polarization fraction and the angle the line of sight makes with the tangential magnetic field lines. However, we do not find evidence for this in sightlines where the dust extinction is dominated by the \lb. We also do not find a correlation between the dust polarization fraction and the inclination angle between the \lb\ wall defined by the model of \citet{O'Neill:2024} and the plane of the sky. Therefore, we conclude that at least one of the commonly made assumptions must not hold. We hypothesize that dust structure beyond the \lb\ wall plays a substantial role in determining the polarization structure of the dust emission. Our results show that simply projecting the Planck polarization data onto the \lb\ geometry is not a well-motivated model for the magnetic field structure of the \lb.

In \secref{sec:3ddust}, we test how dust complexity, i.e., how the~3D dust is distributed along the line of sight, affects the dust polarization fraction. We quantify the dust complexity for each sightline and group the sightlines into low- and high-complexity groups. We pair-match the sightlines across the two groups based on their column densities. For each pair, we subtract the dust polarization fraction of the sightline with low complexity from that of the sightline with high complexity. We find that on average, the dust polarization fraction of the sightlines with higher complexity is~2\% lower than those with lower complexity. This is only true when considering dust out to~1.25~kpc. The result is not statistically different from null when considering dust out to~270~pc only. Note that our definition of complexity does not take into account the distance to different dust components. Future work could incorporate the effect of this distance into the analysis.

We test whether the order of magnitude of this result agrees with our expectation based on geometric depolarization. \citet{Padoan:2001} model the polarized thermal dust emission from protostellar cores formed through supersonic turbulent flows within molecular clouds following the formalism in \citet{Fiege:2000}. \citet{Fiege:2000} develops this formalism to model the submillimeter polarization patterns for filamentary molecular clouds. \citet{Fiege:2000} and \citet{Padoan:2001} ignore the effects of self-absorption and scattering since this model is for submillimeter wavelengths at which the diffuse interstellar medium is optically thin. \citet{Padoan:2001} further assume that the dust grain properties are constant and the temperature is uniform. Since these assumptions are valid for our order of magnitude estimation, we follow the same formalism here.

For this test, we consider sightlines with 2 clouds but different~$\mathcal{N}_c$ values. For simplicity, we assume each cloud to have a constant volume density, plane-of-sky magnetic field angle, and magnetic inclination angle along a certain sightline. Therefore, for a given sightline, we write Equations~\ref{eq:I}, ~\ref{eq:Q}, and~\ref{eq:U} as
\eq{I~=\epsilon N_{\rm H} - \frac{\alpha\epsilon}{2} \left(N_{\rm H, a}\cos^2{\gamma_{\rm a}} + N_{\rm H, b}\cos^2{\gamma_{\rm b}} - \frac{2}{3} N_{\rm H}\right), \label{eq:I2}}
\eq{Q~=~-\alpha\epsilon(N_{\rm H, a}\cos{2\psi_a}\cos^2{\gamma_{\rm a}} + N_{\rm H, b}\cos{2\psi_b}\cos^2{\gamma_{\rm b}})
, \label{eq:Q2}}
\eq{U~=~-\alpha\epsilon(N_{\rm H, a}\sin{2\psi_a}\cos^2{\gamma_{\rm a}} + N_{\rm H, b}\sin{2\psi_b}\cos^2{\gamma_{\rm b}}), \label{eq:U2}}
where~$N_{\rm H, a}$ and~$N_{\rm H, b}$ are the column densities for each cloud,~$N_{\rm H} = N_{\rm H, a} + N_{\rm H, b}$,~$\gamma_{\rm a}$ and~$\gamma_{\rm b}$ are the magnetic inclination angles of the two clouds, and~$\epsilon$ cancels out when calculating~$p$.

We take~$N_{\rm H, a} > N_{\rm H, b}$, i.e.,
\eq{\mathcal{N}_c = \frac{N_{\rm H, a}+N_{\rm H, b}}{N_{\rm H, a}}.}
We divide Equations~\ref{eq:I2}, ~\ref{eq:Q2}, and~\ref{eq:U2} by~$N_{\rm H, a}$ to be able to write them in terms of~$\mathcal{N}_c$. Therefore,
\eq{I/N_{\rm H, a}~=\epsilon\mathcal{N}_c - \frac{\alpha\epsilon}{2} \left[\cos^2{\gamma_{\rm a}} + (\mathcal{N}_c-1)\cos^2{\gamma_{\rm b}} - \frac{2}{3} \mathcal{N}_c\right], \label{eq:I3}}
\eq{Q/N_{\rm H, a}~=-\alpha\epsilon\left[\cos{2\psi_a}\cos^2{\gamma_{\rm a}} + (\mathcal{N}_c-1)\cos{2\psi_b}\cos^2{\gamma_{\rm b}}\right]
, \label{eq:Q3}}
\eq{U/N_{\rm H, a}~=-\alpha\epsilon\left[\sin{2\psi_a}\cos^2{\gamma_{\rm a}} + (\mathcal{N}_c-1)\sin{2\psi_b}\cos^2{\gamma_{\rm b}}\right], \label{eq:U3}}
where~$N_{\rm H, a}$ cancels out when calculating~$p$.

We uniformly sample orientations in the range~$[0,~\pi]$ for~$\psi_a$ and~$\psi_b$, values in the range~$[0,~1]$ for~$\cos{\gamma_{\rm a}}$ and~$\cos{\gamma_{\rm b}}$, and a value in the range~[1,~1.1] for~$\mathcal{N}_c$ to calculate an instance of~$p(\mathcal{N}_c \leq 1.1)$. We also sample different orientations and a value in the range~[1.5,~2] for~$\mathcal{N}_c$ to calculate an instance of~$p(\mathcal{N}_c \geq 1.5)$. We then subtract~$p(\mathcal{N}_c \leq 1.1)$ from~$p(\mathcal{N}_c \geq 1.5)$ as in \secref{sec:3ddust}. We repeat this~10,000 times and average the results. We set~$\alpha~=~0.22$, which corresponds to a maximum dust polarization fraction across the sky~$p_{\rm max}~=~0.22$. This is the value \citet{PlanckCollaboration:2020} observe for~$p_{\rm max}$ at~353~GHz and~80$'$ resolution. We get \eq{\langle p(\mathcal{N}_c~\geq~1.5)~-~p(\mathcal{N}_c~\leq~1.1)\rangle~=~-0.013.} 
This mean difference depends on the value for~$\alpha$. For instance, if we set~$\alpha~=~0.15$ as in \citet{Padoan:2001} instead, we get a mean difference of~-0.009. However, our estimate for the mean difference agrees with the result we measure in \secref{sec:3ddust} for all reasonable values of~$\alpha$. Thus, our empirical result is consistent with our theoretical estimate for the dust depolarization attributable to the line-of-sight dust complexity.

Variations in the orientation of magnetic fields along the line of sight induce differences in the polarization angles of different dust components along the same sightline \citep{Lee:1985,Tassis:2015,King:2018}. When the emission of those components has different spectral energy distributions (SEDs), a frequency-dependent variation of the observed dust polarization angle along that sightline emerges, a phenomenon known as line-of-sight frequency decorrelation. This decorrelation complicates the translation of polarized dust emission maps from one frequency to another. Current analysis within the BICEP/Keck field does not demonstrate evidence of this phenomenon \citep{Ade:2021,Ade:2023}. However, a statistically significant detection of line-of-sight frequency decorrelation has been identified in larger sky areas across sightlines intersecting multiple dust clouds with varying magnetic field orientations \citep{Pelgrims:2021}. Since polarized dust emission is the major foreground for CMB polarization measurements at high frequencies, it is important to characterize how the spatial complexity of the magnetic field in the dust might affect the frequency dependence of the foreground signal. In this paper, we have presented evidence that the~3D spatial complexity of the dust affects the dust polarization signal, even at a fixed frequency.

The analysis in this paper highlights the importance of~3D dust mapping out to large distances. The dust distribution affects the dust polarization fraction, which has implications for the~3D magnetic field distribution. Since we expect higher complexity sightlines to have a lower dust polarization fraction on average, we would infer that the magnetic field is more uniform along a sightline if it has both a highly complex dust distribution and a large dust polarization fraction. This analysis was performed on the sightlines shown in~\figref{fig:masks_complexity}. Improvements to~3D dust modeling will allow us to look for this effect at the very high Galactic latitudes excluded here.

Given the significance of the~3D dust distribution on measurements of the polarized dust emission, these data can be combined with position-position-velocity maps of the neutral hydrogen-based dust polarization templates. These templates, constructed based on the orientation of neutral hydrogen filaments, have been shown to correlate very well with the measured dust polarization \citep{Clark:2019, Cukierman:2023, Ade:2023, Halal:2024}. Since the neutral hydrogen and dust trace similar volumes of the diffuse interstellar medium \citep{Boulanger:1996, Lenz:2017}, future work could morphologically match the position-position-position dust maps with the position-position-velocity neutral hydrogen-based maps to form~4D position-position-position-velocity maps of the magnetic field and polarized dust emission. Starlight polarization can also be used to provide a tomographic view of the plane-of-the-sky magnetic field and polarized dust emission for sightlines with these measurements \citep{Panopoulou:2019,Tassis:2018,Pelgrims:2023,Pelgrims:2024}. These data combined with Faraday tomography \citep{VanEck:2017} or rotation measures~\citep{Tahani:2018} can be used to constrain the~3D magnetic field structure.

\begin{acknowledgments}
We thank Gordian Edenhofer and Minjie Lei for insightful discussions.
This work was supported by the National Science Foundation under grant No. AST-2106607. MT is supported by the Banting Fellowship (Natural Sciences and Engineering Research Council Canada) hosted at Stanford University and the Kavli Institute for Particle Astrophysics and Cosmology (KIPAC) Fellowship.
This publication utilizes data from Planck, an ESA science mission funded by ESA Member States, NASA, and Canada.
The computations in this paper were run on the Sherlock cluster, supported by the Stanford Research Computing Center at Stanford University.
\end{acknowledgments}

%

\vspace{5mm}


\software{astropy \citep{2013A&A...558A..33A,2018AJ....156..123A},
          HEALPix\footnote{\url{http://healpix.sourceforge.net/}} \citep{2005ApJ...622..759G},
          healpy \citep{2019JOSS....4.1298Z},
          matplotlib \citep{2007CSE.....9...90H},
          numpy \citep{10.5555/2886196},
          scipy \citep{2020SciPy-NMeth},
          dustmaps \citep{Green2018},
          astrodendro \citep{Rosolowsky:2008}
          }



\appendix

\section{Analysis Variations} \label{sec:app}

\begin{table*}
\centering 
\begin{tabular}{|c|c|c|c|c|c|c|c|}
\hline
$\Delta_n$ & $\#_{\rm voxels}$ & low complexity & high complexity & $\#$ of matched pairs & $\mu_{\Delta p_{353}}$ & $\sigma_{\Delta p_{353}}$ & p-value\\
\hline
\hline
$\mathbf{7\times10^{-6}}$ & \textbf{3} & $\mathbf{\mathcal{N}_c \leq 1.1}$ & $\mathbf{\mathcal{N}_c \geq 1.5}$ & \textbf{753} & $\mathbf{-1.47\times10^{-2}}$ & $\mathbf{0.22\times10^{-2}}$ & $\mathbf{<0.001}$\\
\hline
\hline
$7\times10^{-6}$ & 3 & $\mathcal{N}_c \leq 1.4$ & $\mathcal{N}_c \geq 2.3$ & 4,674 & $-1.10\times10^{-2}$ & $0.08\times10^{-2}$ & $<0.001$\\
\hline
$9\times10^{-6}$ & 3 & $\mathcal{N}_c \leq 1.1$ & $\mathcal{N}_c \geq 1.5$ & 776 & $-1.33\times10^{-2}$ & $0.26\times10^{-2}$ & $<0.001$\\
\hline
$9\times10^{-6}$ & 3 & $\mathcal{N}_c \leq 1.4$ & $\mathcal{N}_c \geq 2.3$ & 4,719 & $-1.01\times10^{-2}$ & $0.12\times10^{-2}$ & $<0.001$\\
\hline

$5\times10^{-6}$ & 5 & $\mathcal{N}_c \leq 1.1$ & $\mathcal{N}_c \geq 1.5$ & 1,005 & $-0.98\times10^{-2}$ & $0.20\times10^{-2}$ & $<0.001$\\
\hline
$5\times10^{-6}$ & 5 & $\mathcal{N}_c \leq 1.4$ & $\mathcal{N}_c \geq 2.3$ & 5,311 & $-0.82\times10^{-2}$ & $0.14\times10^{-2}$ & $<0.001$\\
\hline

$3\times10^{-6}$ & 5 & $\mathcal{N}_c \leq 1.1$ & $\mathcal{N}_c \geq 1.5$ & 984 & $-1.00\times10^{-2}$ & $0.21\times10^{-2}$ & $<0.001$\\
\hline
$3\times10^{-6}$ & 5 & $\mathcal{N}_c \leq 1.4$ & $\mathcal{N}_c \geq 2.3$ & 5,284 & $-0.80\times10^{-2}$ & $0.13\times10^{-2}$ & $<0.001$\\
\hline

\end{tabular}
\caption{Results obtained from varying both the dendrogram parameters~($\Delta_n$ and $\#_{\rm voxels}$) used to identify the density peaks in the line-of-sight dust distribution and from varying the upper and lower thresholds in~$\mathcal{N}_c$ used to divide the sightlines into low- and high-complexity groups. The results of the main analysis are bolded in the first row. The mean and standard deviation in the~$\mu_{\Delta p_{353}}$ and~$\sigma_{\Delta p_{353}}$ columns are over the~12 posterior samples of the \citet{Edenhofer:2023} maps.
\label{tab:params}}
\end{table*}

\begin{table*}
\centering 
\begin{tabular}{|c|c|c|c|c|c|c|c|}
\hline
$\Delta_n$ & $\#_{\rm voxels}$ & low complexity & high complexity & $\#$ of matched pairs & $\mu_{\Delta p_{353}}$ & $\sigma_{\Delta p_{353}}$ & p-value\\
\hline
$7\times10^{-6}$ & 3 & $\mathcal{N}_c \leq 1.1$ & $\mathcal{N}_c \geq 1.5$ & 753 & $-1.45\times10^{-2}$ & $0.23\times10^{-2}$ & $<0.001$\\
\hline
$7\times10^{-6}$ & 3 & $\mathcal{N}_c \leq 1.4$ & $\mathcal{N}_c \geq 2.3$ & 4,674 & $-1.02\times10^{-2}$ & $0.08\times10^{-2}$ & $<0.001$\\
\hline

$9\times10^{-6}$ & 3 & $\mathcal{N}_c \leq 1.1$ & $\mathcal{N}_c \geq 1.5$ & 776 & $-1.31\times10^{-2}$ & $0.34\times10^{-2}$ & $<0.001$\\
\hline
$9\times10^{-6}$ & 3 & $\mathcal{N}_c \leq 1.4$ & $\mathcal{N}_c \geq 2.3$ & 4,719 & $-1.00\times10^{-2}$ & $0.15\times10^{-2}$ & $<0.001$\\
\hline

$5\times10^{-6}$ & 5 & $\mathcal{N}_c \leq 1.1$ & $\mathcal{N}_c \geq 1.5$ & 1,005 & $-0.60\times10^{-2}$ & $0.17\times10^{-2}$ & $<0.001$\\
\hline
$5\times10^{-6}$ & 5 & $\mathcal{N}_c \leq 1.4$ & $\mathcal{N}_c \geq 2.3$ & 5,311 & $-0.82\times10^{-2}$ & $0.13\times10^{-2}$ & $<0.001$\\
\hline

$3\times10^{-6}$ & 5 & $\mathcal{N}_c \leq 1.1$ & $\mathcal{N}_c \geq 1.5$ & 984 & $-0.67\times10^{-2}$ & $0.22\times10^{-2}$ & $<0.001$\\
\hline
$3\times10^{-6}$ & 5 & $\mathcal{N}_c \leq 1.4$ & $\mathcal{N}_c \geq 2.3$ & 5,284 & $-0.80\times10^{-2}$ & $0.13\times10^{-2}$ & $<0.001$\\
\hline

\end{tabular}
\caption{Results obtained from the same variations described in \tabref{tab:params}, where the sightline matching over the two complexity groups  in this case includes the absolute value of the Galactic latitude as an additional confounding variable. The mean and standard deviation in the~$\mu_{\Delta p_{353}}$ and~$\sigma_{\Delta p_{353}}$ columns are over the~12 posterior samples of the \citet{Edenhofer:2023} maps.
\label{tab:match}}
\end{table*}

\begin{table}
\centering 
\begin{tabular}{|c|c|c|c|c|}
\hline
$\Delta_n$ & $\#_{\rm voxels}$ & $\mu_{\Delta p_{353}}$ & $\sigma_{\Delta p_{353}}$ & p-value\\
\hline
$7\times10^{-6}$ & 3 & $-1.08\times10^{-2}$ & $0.17\times10^{-2}$ & $<0.001$\\
\hline

$9\times10^{-6}$ & 3 & $-1.14\times10^{-2}$ & $0.24\times10^{-2}$ & $<0.001$\\
\hline

$5\times10^{-6}$ & 5 & $-0.84\times10^{-2}$ & $0.17\times10^{-2}$ & $<0.001$\\
\hline

$3\times10^{-6}$ & 5 & $-0.89\times10^{-2}$ & $0.18\times10^{-2}$ & $<0.001$\\
\hline

\end{tabular}
\caption{Results obtained from sampling~1,000 sightlines from each of the~$\mathcal{N}_c~\leq~1.4$ and~$\mathcal{N}_c~\geq~2.3$ complexity groups before matching. The different rows are for different variations in the dendrogram parameters~($\Delta_n$ and $\#_{\rm voxels}$) used in identifying the density peaks in the line-of-sight dust distribution. The mean and standard deviation in the~$\mu_{\Delta p_{353}}$ and~$\sigma_{\Delta p_{353}}$ columns are over the~12 posterior samples of the \citet{Edenhofer:2023} maps.
\label{tab:samp}}
\end{table}

In this appendix, we list some of the results obtained from varying the main analysis choices in \secref{sec:3ddust}. These analysis variations are described and their results summarized in \secref{subsec:valid}. In \tabref{tab:params}, we present some of the results obtained from varying the dendrogram parameters used in identifying the density peaks in the line-of-sight dust distribution and from varying the upper and lower thresholds in~$\mathcal{N}_c$ used for splitting the sightlines into low- and high-complexity groups. \tabref{tab:match} lists the results of the same variations performed in \tabref{tab:params} but when including the absolute value of the Galactic latitude as an additional confounding variable in the sightline matching across the two complexity groups. Finally, we present the results of sampling~1,000 sightlines from each of the two complexity groups with~$\mathcal{N}_c\leq1.4$ and~$\mathcal{N}_c\leq2.3$ before matching in \tabref{tab:samp}. We find that the result in our main analysis is robust to all of these analysis variations.


\bibliography{LB}{}
\bibliographystyle{aasjournal}



\end{document}